\documentclass[usegraphix,usenatbib]{mn2e}

\newcommand{\kpc}{\rm\thinspace kpc}
\newcommand{\pc}{\rm\thinspace pc}

\usepackage{amsmath}
\usepackage{amssymb}
\usepackage{times}
\usepackage{graphicx}
\usepackage{subfigure}
\usepackage{epsfig}
\usepackage{txfonts}
\usepackage{epstopdf}

\def\intd{{\mathrm{d}}}
\def\del{{\vec{\nabla}}}
\def\acts{{\boldsymbol{J}}}
\def\pos{{\boldsymbol{x}}}
\def\vel{{\boldsymbol{v}}}

\def\kpc{{\mathrm{kpc}}}
\def\pc{{\mathrm{pc}}}
\def\Ms{{\mathrm{M_\odot}}}

\begin{document}

\pagenumbering{arabic}

\title[Galaxy Models in Action Coordinates]
  {Made--to--Measure Dark Matter Haloes, Elliptical Galaxies and Dwarf Galaxies in Action Coordinates}

\author[Williams \& Evans]
  {A.A. Williams$^1$\thanks{E-mail: aamw3,nwe@ast.cam.ac.uk},
   N.W. Evans$^1$
 \medskip
 \\$^1$Institute of Astronomy, University of Cambridge, Madingley Road,
       Cambridge, CB3 0HA, UK}

\maketitle

\begin{abstract}
We provide a family of action-based distribution functions (DFs) for
the double--power law family of densities often used to model
galaxies. The DF itself is a double--power law in combinations of the
actions, and reduces to the pure
power--law case at small and large radii.  Our method enables the velocity
anisotropy of the model to be tuned, and so the anisotropy in the
inner and outer parts can be specified for the application in hand. We
provide self-consistent DFs for the Hernquist and Jaffe models -- both
with everywhere isotropic velocity dispersions, and with kinematics
that gradually become more radially anisotropic on moving
outwards. We also carry out this exercise for a cored dark--matter model. 
These are tailored to represent dark haloes and elliptical
galaxies respectively with kinematic properties inferred from
simulations or observational data. Finally, we relax a cored luminous
component within a dark matter halo to provide a self-consistent model
of a dwarf spheroidal embedded in dark matter. The DFs provide us with
non-rotating spherical stellar systems, but one of the virtues of
working with actions is the relative ease with which such models can
be converted into axisymmetry and triaxiality.
\end{abstract}

\begin{keywords}
methods: analytical - Galaxy: kinematics and dynamics - galaxies: kinematics and dynamics
\end{keywords}

\section{INTRODUCTION}
The interpretation of kinematic observations of galaxies in terms of
their orbital structure is an important and difficult problem in
modern-day galaxy dynamics.  For external galaxies, the kinematics may
include mean motions, velocity dispersions and even line profiles, or
the entire distribution of line of sight velocities.  For the Milky
Way, the problem has been given additional impetus by the advent of
large-scale spectroscopic surveys, together with the launch of the
Gaia satellite, which will provide discrete velocities on about a
billion objects (e.g., \citealt{Pe01}). Moment based methods, such as
the Jeans equations, offer a simple and widely used method for
reproducing the density and velocity dispersion (e.g.,
\citealt{Fi86}).  Nonetheless, they are much less powerful than
distribution function (or DF) based methods, which can directly fit
not just mere moments, but also the distributions of kinematical
quantities (e.g., \citealt{Wo10,Am11,Wo13a,St14}).

However, construction of equilibrium DFs for galaxies is far from
easy. There do exist numerical algorithms such as Schwarzschild's
(1979) \nocite{Sc79} orbit superposition method or Syer \& Tremaine's
(1996) \nocite{Sy96} made-to-measure method. These can be thought of
as ways to fit orbits or N-body models to kinematic data, such that
the deviation of moments from the observables are minimized subject to
some penalty function that enforces smoothness. Schwarzschild's
method, at least in its axisymmetric implementation, has proved
invaluable in the analysis of integral-field stellar kinematics on
elliptical galaxies (e.g., \citealt{Ca07}). The made-to-measure method
has also been applied to elliptical galaxies to assess their dark
matter content and orbital anisotropy \citep{De08}.

There also exist classes of simple analytic DFs, though these are
restricted to spherical symmetry or to axisymmetry (e.g.,
\citealt{De86}).  DFs can only depend on the isolating integrals of
motion by \citet{Je19} theorem. In spherical potentials, they are the
energy $E$ and angular momentum components $\bmath{L}$; in
axisymmetric potentials, they are energy $E$ and component of angular
momentum parallel to the symmetry axis $L_z$. Given the density,
general algorithms exists to find smooth DFs based on the classical
integrals (\citealt{Ed15}, \citealt{Hu93}, \citealt{Ev06}). There also
systems with known DFs, including power-laws \citep{Ev94} and double
power-laws, such as \citet{He90} and \citet{Ja83} models.

Binney (2008) \nocite{Bi08} has argued that it is more natural to use
actions as the choice of integrals of motion rather than classical
integrals, such as energy.  This is partly because the actions are
adiabatic invariants, and partly because action-based DFs can be more
easily generalized to flattened and triaxial geometries.  Binney
(2010, 2012) \nocite{Bi10,Bi12b} provided a significant advance when
he showed that the data on the thin disk of the Milky Way can be
largely accounted for by models synthesised from quasi-isothermal
DFs. The thin disk provides a particularly clean application for two
reasons. First, the stellar orbits are close to circular and so the
actions of stars are readily computed in terms of epicyclic theory and
its extensions. Second, the quasi-isothermal assumption provides a
simple and physically motivated ansatz for the DF, building on earlier
ideas that the DF is Maxwellian about the circular speed \citep{Sh67}.

The purpose of this paper is to extend this work to hotter, or
pressure-supported, stellar systems. That is, we seek similar
action-based DFs for elliptical galaxies or bulges or haloes. Ideally,
the DFs should be tunable, so that the user may specify the power-law
fall-off in the density at large and small radii, as well as
properties of the velocity anisotropy at large and small radii. Such
an algorithm provides the user with a way of making made-to-measure
haloes or elliptical galaxies.

\section{Background}

Here, we introduce some important concepts relating to this work that
will prove important in the main body of the paper. First, we describe
how to compute a self--consistent model given an action--based
distribution function. Then,we describe how to construct constant
anisotropy DFs for simple power--law models.

\subsection{Computing a Self--Consistent Model}
\label{sec:selfcons}

Action integrals are adiabatically invariant, which means that slow
changes in $\Phi(r)$ do not alter $\acts$. By extension, an
action--based distribution function $f(\acts)$ is also adiabatically
invariant. This allows us to propose a model $f(\acts)$ and
iteratively converge upon its corresponding potential, $\Phi _f$, from
an initial educated guess $\Phi _0$. In a spherical system, the
relevant actions are the azimuthal action $J_\phi = L_z$, the vertical action 
$J_\theta = L - |L_z|$, and the radial action
\begin{equation} \label{eq:gensol}
J_{r} = \dfrac{1}{2\pi}\oint\sqrt{-2E-2\Phi(r)-L^{2}/r^{2}}\mathrm{d}r.
\end{equation}
However, for a non--rotating spherical system, one can show that the Hamiltonian (and hence the DF) can be written as a function of 
just $L = J_\theta + |J_\phi|$ and $J_r$. We thus write $f(\acts) = f(L,J_r)$. The algorithm to find the self--consistent
model is as follows \citep{Bi08}
\begin{enumerate}
\item From the initial guess potential, $\Phi _0$, compute the radial
  action $J_r(\pos,\vel)$ as a function of the phase--space
  coordinates (the other action, $L$, is potential--independent).
\item Compute the implied density profile of $f(\acts)$ under $\Phi
  _0$ by carrying out the integral
\begin{equation}
\rho_1(\pos) = \int \intd ^3 \vel f\left(L(\pos,\vel),J_r(\pos,\vel)\right).
\end{equation}
\item Solve Poisson's equation to find the potential implied by $\rho
  _1$
\begin{equation}
\del ^2 \Phi _1(\pos) = 4\pi G \rho _1(\pos).
\end{equation}
\item Repeat steps (i) $\rightarrow$ (iii) with $\Phi _1$ in place of
  $\Phi _0$ and compute $\Phi _2$.
\item Iterate until the difference between $\Phi _i$ and $\Phi _{i+1}$
  is negligible.
\end{enumerate}
Once the algorithm has converged, and we possess $\Phi _f$, we can
compute the radial action (and therefore DF) as a function of the
phase--space coordinates $(\pos,\vel)$. This means that we can compute
any observable we choose in a self--consistent fashion. The adiabatic
invariance of $f(\acts)$ means that we can create models composed of
many different components (e.g. galaxy models with a dark halo, disk
and bulge) and relax them simultaneously. It is this unique feature of
$f(\acts)$ models that makes them so useful.

\begin{figure}
\includegraphics[width=3.4in]{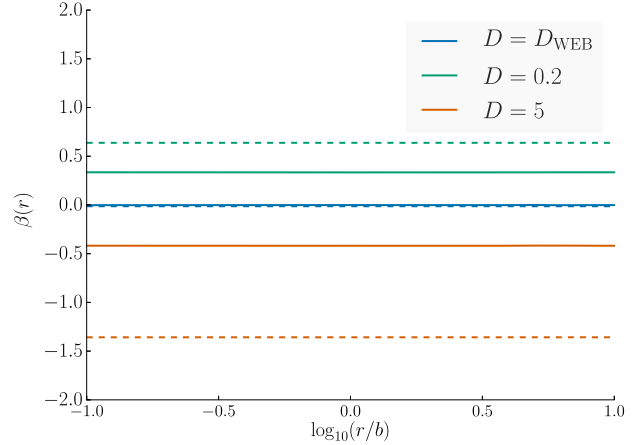}
\caption{Anisotropy profiles for potentials with $\Phi \propto
  r^{\pm 1/2}$. Solid lines correspond to models where $\epsilon = 1/2$ and dashed lines to $\epsilon = -1/2$. $D_{\mathrm{WEB}}$ is obtained from Equation (\ref{eq:consts}), giving very nearly
  isotropic models, $D=0.2$ creates radially biased models and $D=5$ creates tangentially biased models. All the models have constant anisotropy, as
  predicted.}
\label{fig:powlawbeta}
\end{figure}

\subsection{Power--Law Models}
\label{sec:powerlaws}
Before we derive DFs for more complex models, it is instructive to
consider the simpler task of constructing distribution functions for a
power--law model with density law:
\begin{equation}
\rho (r) = \rho _0 \left(\dfrac{r}{a}\right)^{-\nu}.
\label{eq:powdens}
\end{equation}
Such a density distribution will generate a gravitational potential
$\Phi(r) \propto r^\epsilon$, where $\epsilon = 2- \nu$. We can then
construct isotropic distribution functions for these systems as a
function of binding energy \citep{Ev94}
\begin{equation}
f(E) \propto |E|^{-\dfrac{\epsilon+4}{2\epsilon}}.
\label{eq:powlaw}
\end{equation}
An obvious way to obtain a distribution function in action space is to
then express the binding energy (given by $E=-H$, where $H$ is the
Hamiltonian) as a function of the actions $\acts$, and substitute the
resulting expression into Equation (\ref{eq:powlaw}). \citet*{Wi14a}
(hereafter WEB) provide an approximation to the Hamiltonians of
power--law models, given by
\begin{equation}
\mathcal{H}(\acts) \propto (L + DJ_{r})^{\zeta},
\label{eq:approxham}
\end{equation}
where $\zeta = 2\epsilon/(\epsilon +2)$ and
\begin{equation} \label{eq:consts}
D = \begin{cases}
\dfrac{\sqrt{2\pi}\Gamma(3/2+ 1/\epsilon)\epsilon^{-1/\epsilon}\zeta^{1/\zeta}}{\Gamma(1+1/\epsilon)} & \epsilon > 0 \\
\sqrt{2\mathrm{\pi}/e} & \epsilon = 0 \vspace{6pt}\\
\dfrac{\sqrt{2\pi}\Gamma(1-1/\epsilon)(-\epsilon)^{1-1/\epsilon}(-\zeta)^{1/\zeta}}{\Gamma(-1/\epsilon-1/2)} & \epsilon < 0. \\ \end{cases} 
\end{equation}
Given Equation (\ref{eq:approxham}), the distribution function in
action--space is given by
\begin{equation}
f(\acts) \propto (L + DJ_{r})^{-(\epsilon+4)/(\epsilon+2)}.
\label{eq:powdf}
\end{equation}
This expression is therefore an approximate isotropic distribution
function for the power law in Equation (\ref{eq:powdens}) when
$0\leq\nu < 3$. We note that Equation (\ref{eq:powdf}) still holds in
the case $\nu=2$ and the potential is logarithmic. We now turn to the
problem of constructing constant anisotropy power--law models. The
anisotropy parameter is given by
\begin{equation}
\beta(r) = 1 - \dfrac{\sigma ^2_t(r)}{2\sigma ^2 _r(r)},
\end{equation}
where $\sigma _r$/$\sigma _t$ are the radial/tangential velocity
dispersions. $\beta$ quantifies the relative importance of radial and
tangential orbits at radius $r$: when a model is completely
constructed from radial/circular orbits $\beta \rightarrow 1 /
-\infty$. To construct constant anisotropy power--laws, we can take
the commonly used ansatz (e.g., Wilkinson \& Evans 1999, Evans \& An
2006, Deason, Belokurov \& Evans 2011) \nocite{Wi99,Ev06,De11a}
\begin{equation}
f(E,L) = L^{-2\beta}g(E),
\end{equation}
and once again substitute for $E$ using Equation
(\ref{eq:approxham}). However, we can also note that a density profile
with the same radial dependence as Equation (\ref{eq:powdens}) is
generated by \textit{any} non--negative, scale--free DF with the
dimensions of Equation (\ref{eq:powdf}). As a consequence, we can
construct a family of constant anisotropy power laws using a DF of the
form of Equation (\ref{eq:powdf}). When $\nu$ is fixed, the isotropic
model belonging to this family is generated when the parameter $D$ is
set equal to the value implied by Equation (\ref{eq:consts}), and
anisotropic models can be generated by altering the value of this
parameter. The anisotropy profile must be constant with radius for all
such models, as they possess no scale. We can intuitively understand
how different values of $D$ will produce different anisotropies. $D$
controls the relative importance of tangential and radial orbits: a
model that more heavily weights the angular momentum will become
tangentially biased, whereas favouring the radial action will result
in radial bias. Since the DF is a declining function of $\acts$, we
can expect that an increase in $D$ from the WEB value of Equation
(\ref{eq:consts}) will produce a tangentially biased model, and a
decrease will result in a radially biased model.

As an example, Figure \ref{fig:powlawbeta} depicts the variation in the anisotropy parameter 
for three choices of $D$ in the cases $\epsilon = \pm 1/2$. The three different values are
$D_{\mathrm{WEB}}$ from Equation (\ref{eq:consts}), 0.2 and 5. $D_{\mathrm{WEB}}$ produces very nearly 
isotropic models, $D=5$ creates tangentially biased models and $D=0.2$
creates radially biased models. All the models have constant anisotropy as expected.    

\begin{figure*}
\begin{centering}
\includegraphics[width=3in]{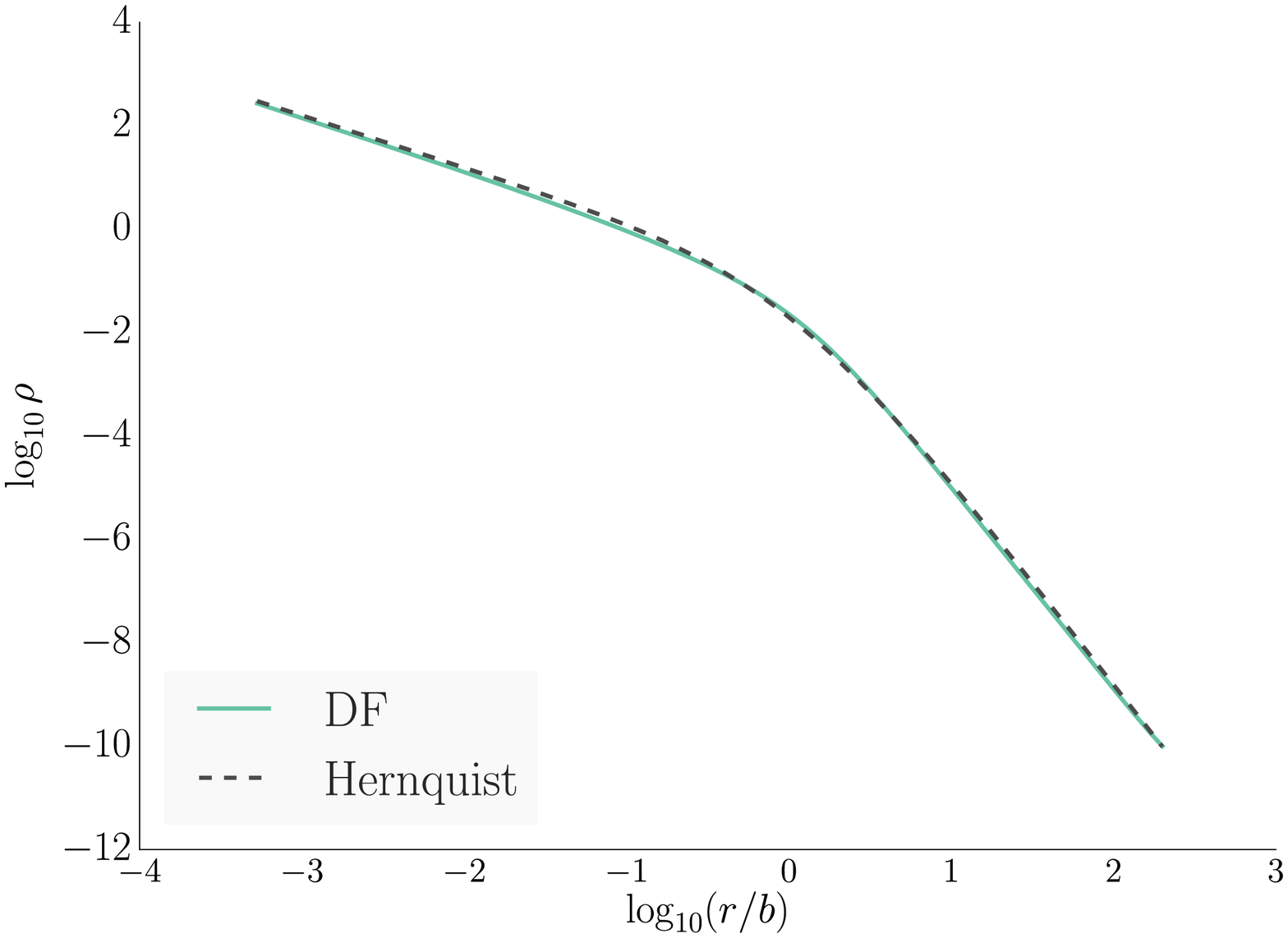}\quad\includegraphics[width=3in]{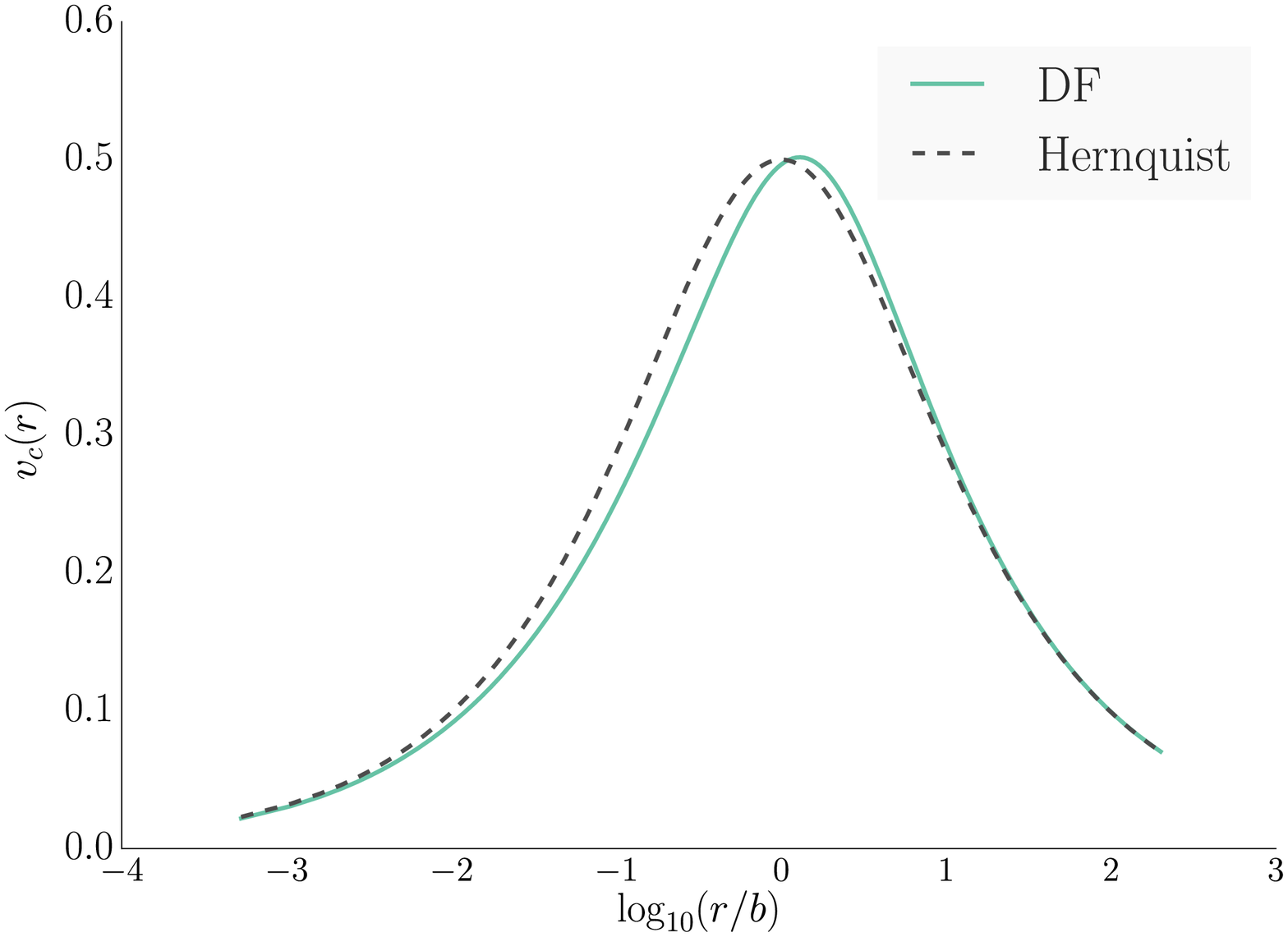} \\
\includegraphics[width=3in]{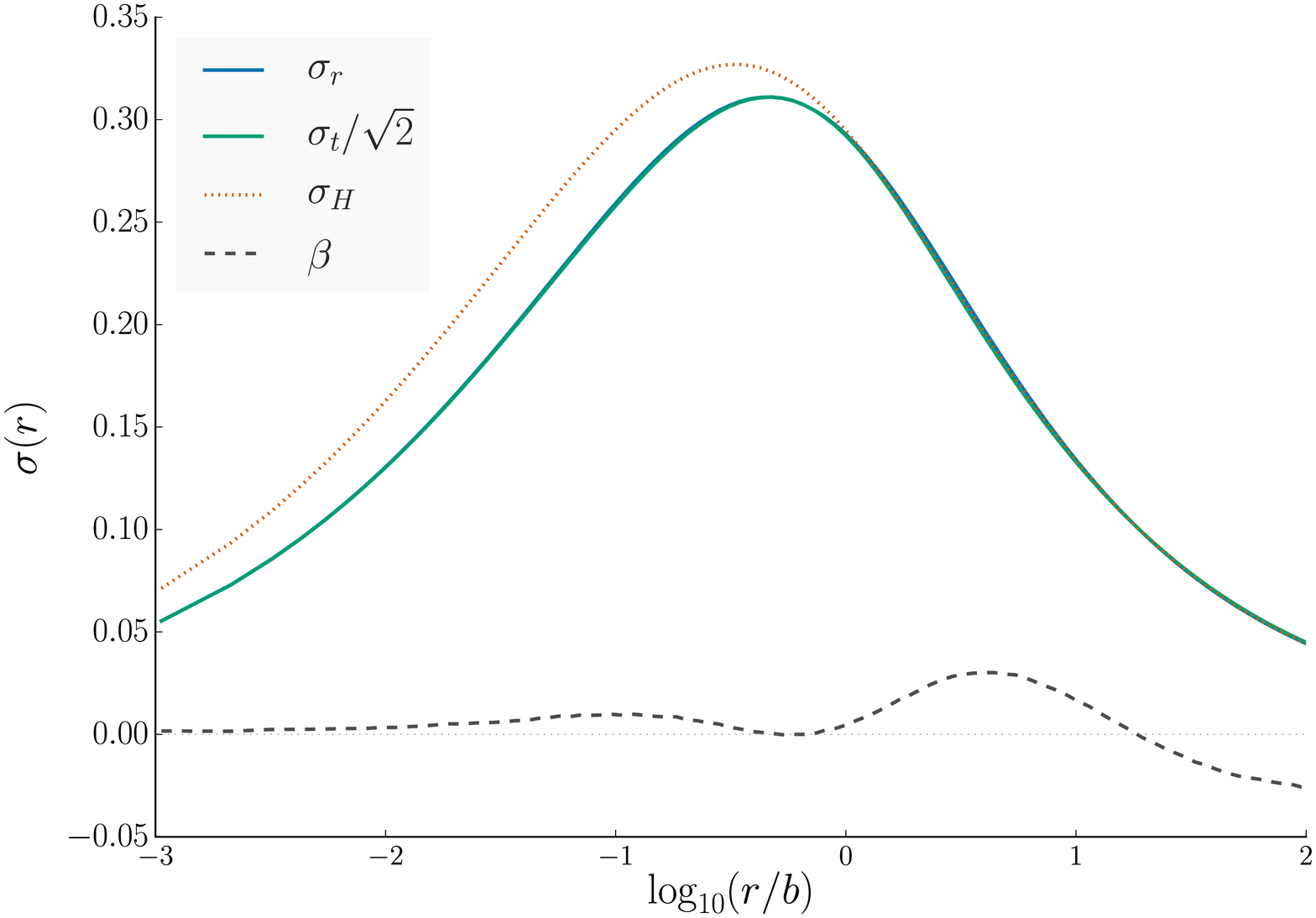}\quad\includegraphics[width=3in]{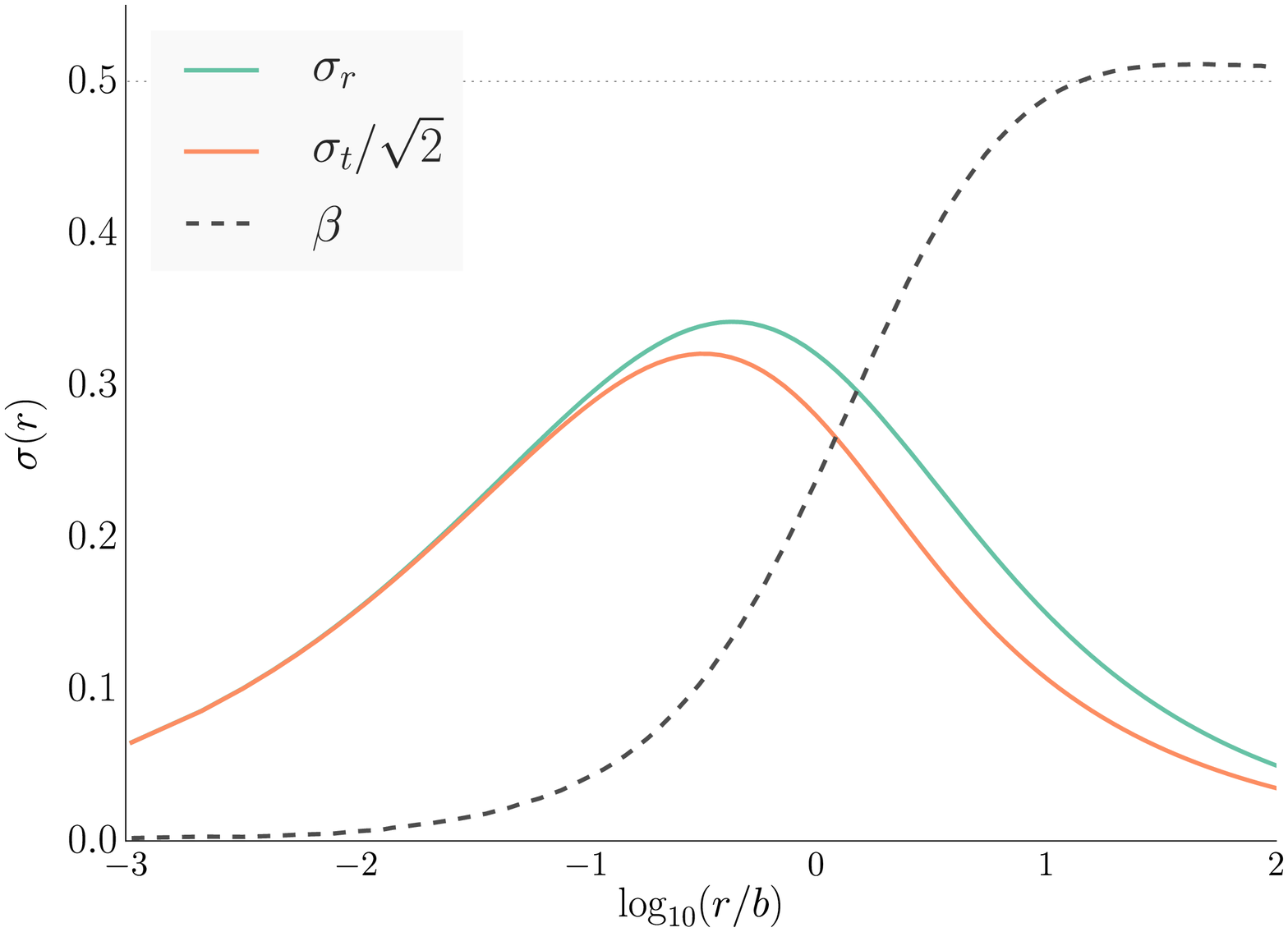}
\caption{Top row: comparison of the density (left) and circular speed
  (right) profiles generated by the DF (\ref{eq:hernDF}) and the
  Hernquist model. The densities are in excellent agreement, and the
  circular speed curves match well, other than a mild offset visible
  at small radii. Bottom row: the kinematics of the two dark matter
  models. The model intended to be isotropic (left) is almost so, with
  $\beta$ only fluctuating on scales of $\sim 0.01$. The dotted line depicts the isotropic 
  velocity dispersion of the equivalent isotropic Hernquist model ($\sigma _H$), which agrees 
  well with our models other than a small offset in the cusp of the model. The radially
  distended model (right) moves smoothly from $\beta _0 = 0$ to $\beta _1 =
  0.5$ across a scale of $r_\beta = b$. We set $G=M=b=1$ for these
  models.}
\label{fig:hernfig}
\par\end{centering}
\end{figure*}

\section{Double--Power law DF}

Having deduced how to produce constant anisotropy power--law
distribution functions, we now aim to extend this reasoning to
construct a family of distribution functions that will generate double
power--law density profiles (see e.g. \citealt{Bi08}):
\begin{equation}
\rho (r) = \dfrac{\rho _0 b^\gamma}{r^\alpha (b+r)^{\gamma-\alpha}}.
\label{eq:doubledens}
\end{equation}
We restrict ourselves to finite--mass models in this case,
i.e. $\gamma > 3$. Such a density profile breaks around the radius
$r=b$, behaving as $r^{-\alpha}$ when $r\ll b$ and as $r^{-\gamma}$
when $r\gg b$. It is by consideration of these power--law limits, in
combination with the reasoning from Section \ref{sec:powerlaws},
that will allow us to construct a suitable DF to emulate the density
profile of Equation (\ref{eq:doubledens}). In what follows, we shall
derive our distribution function designed to mimic these models, then
describe how to tune the anisotropy parameter.

\subsection{Derivation of the DF}

Consider the isotropic distribution function of the double power--law
models, $f(E)$: even if we cannot explicitly calculate its functional
form everywhere, we can infer what it should look like in the limits
of high/low binding energy. At high binding energies, the overwhelming
contribution to the density profile will come from orbits that reside
at radii $r \ll b$. As a result, the DF will resemble that of a
power--law model with density $\rho \sim r^{-\alpha}$, with the same form 
as Equation (\ref{eq:powdf}). 

At low binding energies orbits reside in a Keplerian potential, owing
to the assumed finite mass of the model. In this case, we turn to
Eddington's equation to discover the behaviour of $f(E)$:
\begin{equation}
f(E) \propto \dfrac{\intd}{\intd E} \int ^E _0 \dfrac{\intd\Psi}{\sqrt{E-\Psi}} \dfrac{\intd \rho}{\intd \Psi}, 
\end{equation}
where $\Psi = -\Phi$ is the relative potential. In this case, we have
that $\rho \propto \Psi ^ \gamma$, and Eddington's equation may be
solved to give
\begin{equation}
f(E) \propto E^{\gamma - \frac{3}{2}}.
\end{equation}
An expression for $E$ in terms of $\acts$ is now required. At low binding energies the Hamiltonian
coincides with the Kepler case (e.g., \citealt{Go80})
\begin{equation}
H(\acts) = \dfrac{-(GM)^2}{2\left(L+J_r\right)^2},
\label{eq:kepler}
\end{equation}
where $M$ is the mass of the model. We have thus discovered the limiting behaviour of $f(\acts)$ to be
\begin {equation}
f(\acts) = \begin{cases}
S_\alpha(L+D_\alpha J_r)^{(\alpha-6)/(4-\alpha) } & E \ll -\Phi(b) \\
S_\gamma(L+J_r)^{-2\gamma + 3} & E \gg -\Phi(b), \end{cases}
\label{eq:asymDF}
\end{equation}
where $S_\alpha$ and $S_\gamma$ are constants, and $D_\alpha$ is given by Equation (\ref{eq:consts}) with $\alpha = \nu$. We must now
construct a full DF that satisfies these limits, whilst also
recovering sensible behaviour in--between. Given that the density law
itself is a double power--law, a simple first guess at an appropriate
functional form is a double power--law in the actions. Hence, we
choose the following DF:
\begin{equation}
f(\acts) = \dfrac{\mathcal{N} \, M}{(2\mathrm{\pi})^3J_0 ^{3-\mu}} \,\dfrac{ T(\acts) \, \mathcal{L}(\acts)^{-\lambda} }{\left[J_0^2 + \mathcal{L}(\acts)^2\right]^{(\mu - \lambda)/2}},
\label{eq:crux}
\end{equation}
where we have set $\lambda = (6-\alpha)/(4-\alpha)$, $\mu = 2\gamma - 3$ and 
\begin{eqnarray}
\mathcal{L}(\acts) &=& L + D(\acts)J_r \nonumber \\
D(\acts) &=& \dfrac{D_0 + D_1|\acts|/J_\beta}{1+|\acts|/J_\beta}. \\
T(\acts) &=& \dfrac{S_\alpha+S_\gamma|\acts|/J_0}{1+|\acts|/J_0} \nonumber \\
|\acts| &=& \sqrt{L^2 + J_r^2} \nonumber
\label{eq:defD}
\end{eqnarray}
Several of the components in this DF require explanation. The natural
action scale of the model is
\begin{equation}
J_0 = \sqrt{GMb},
\end{equation}
and controls the transition from one power--law regime to the
other. The argument $\mathcal{L}(\acts)$ is inspired by the linearity
of the Hamiltonian of the model in the two limits. However, the planes
in $\acts$--space upon which the DF is stratified change with growing
$\acts$, as the model passes between the two regimes. This change
is facilitated by the function $D(\acts)$. The number $D_0$/$D_1$
quantifies the slope of these planes at high/low binding energies
(small/large action), and $D(\acts)$ transits $D_0 \rightarrow D_1$
over the action scale $J_\beta$. For the ergodic case, $D_0$ can be
found analytically using Equation (\ref{eq:consts}) and $D_1 = 1$. The
action scale $J_\beta$ differs from the natural action scale, and will
help us to construct models with made--to--measure kinematics. This
will be explained in detail shortly.

In general $f(E)$ is not a double--power law in energy, but is usually
rather more complicated. For some models (e.g. \citealt{He90},
\citealt{De93}, \citealt{Tr94}) $f(E)$ can be computed wholly
analytically, and the added complexity becomes transparent. However,
Equation (\ref{eq:crux}) assumes that $f(E)$ can be approximated by
interpolating between the limiting cases as a double--power law. If we
are to do this effectively, we must ensure that the limiting cases we
sew together have the correct relative normalisation. This amounts to
computing the constant factors $S_\alpha$ and $S_\gamma$ in Equation
(\ref{eq:asymDF}), so that $f(\acts)$ coincides exactly with the
correct limits. $T(\acts)$ fulfills this purpose by acting as a
``variable normalisation" of the DF, interpolating between $S_\alpha$
and $S_\gamma$ over the action scale of the model. For many models
(even if $f(E)$ cannot be represented by elementary functions)
$S_\alpha$ and $S_\gamma$ can be computed analytically. See Appendix A
for a derivation of these quantities in the case $\gamma = 4$.

Finally, the normalisation factor $\mathcal{N}$ ensures that the DF
integrates to the correct mass:
\begin{equation}
M  = (2\pi)^3\int \intd ^3 \acts f(\acts)
\end{equation}
and is computed by a swift numerical integration.

\subsection{Tuning the Anisotropy Profile}
\label{sec:anistune}

Here, we describe an algorithm to tailor the anisotropy of a model. We
can use the logic found in Section \ref{sec:powerlaws} to tune the
anisotropy of our models in the central/far--field regimes.  Using
Equation (\ref{eq:defD}), in the limit $|\acts| \ll J_\beta$, we find
that $D \rightarrow D_0$. In the opposite limit, $D \rightarrow
D_1$. The action scale $J_\beta$ controls the speed at which we
transit from one limit to the other. This allows us to vary $D_0$
until we reach the desired inner anisotropy $\beta _0$, then
independently vary $D_1$ until the outer anisotropy is some value
$\beta _1$. Once the values of $D_0$ and $D_1$ are fixed, we can vary
how fast the model transits from $\beta _0$ to $\beta _1$ by altering
the value of $J_\beta$.

An important subtlety of this procedure is that the relative
normalisation factor $T(\acts)$ must change as a consequence. Consider
the value of the DF at some point in action space $\acts = J_L(1,1)$
where $J_L$ is large. Before the transformation of $D_1$ (we begin
with $D_1 = 1$, the isotropic value), this is equal to
\begin{equation}
f(\acts) = \mathcal{N}S_\gamma (2J_L)^{-\mu}.
\end{equation}
After changing $D_1$, this becomes
\begin{equation}
f'(\acts) = \mathcal{N} S_\gamma \left[(1+D_1)J_L\right]^{-\mu}.
\end{equation}
Thus the weight of the DF at this point in action space has changed by a factor
\begin{equation}
\Delta f = \left(\dfrac{1+D_1}{2}\right)^{-\mu}.
\end{equation}
This suggests we make the transformation
\begin{equation}
S_\gamma \rightarrow S_\gamma \left(\dfrac{2}{1+D_1}\right)^{-\mu}
\label{eq:farfieldnorm}
\end{equation}
in order to preserve the weight of the DF in the far--field. Similar
logic applies in the center of the model, leading to the
transformation
\begin{equation}
S_\alpha \rightarrow S_\alpha \left(\dfrac{1+D_\mathrm{WEB}}{1+D_0}\right)^{-\lambda}
\label{eq:centernorm}
\end{equation}
where $D_\mathrm{WEB}$ is the isotropic value from Equation
(\ref{eq:consts}). After these transformations, the density profile
should barely be altered as a consequence of the anisotropy
tuning. Our algorithm is then as follows:
\begin{enumerate}
\item Choose a central anisotropy, $\beta _0$ and a far--field
  anisotropy $\beta _1$. In addition, specify a length scale $r_\beta$
  over which the anisotropy parameter $\beta(r)$ should transit
  between these values.
\item Compute the self--consistent isotropic model with $D_0$ =
  $D_\mathrm{WEB}$, $D_1 = 1$ and $J_\beta = 1$.
\item Iteratively compute $D_0$ by recalculating
  $\beta(r_\mathrm{inner})$ repeatedly until
  $\beta(r_\mathrm{inner})=\beta _0$, where $r_\mathrm{inner} \ll
  r_\beta$.
\item Iteratively compute $D_1$ by recalculating
  $\beta(r_\mathrm{outer})$ repeatedly until
  $\beta(r_\mathrm{outer})=\beta _1$, where $r_\mathrm{outer} \gg
  r_\beta$.
\item Minimise the function $|\beta(r_\beta) - \frac{1}{2} (\beta _0 +
  \beta _1)|$ at fixed $D_0$, $D_1$ to constrain $J_\beta$.
\item Make the transformations from Equations (\ref{eq:farfieldnorm})
  and (\ref{eq:centernorm}) to minimise changes to the density profile
  of the model.
\end{enumerate}

If $\beta _0 = \beta _1 = 0$ then $D_0$ and $D_1$ are both known
analytically. In this case, only step (v) of the algorithm is
applied, and serves to minimise fluctuations in $\beta(r)$ across all
radii.

\section{Applications}

Here, we use the DF (\ref{eq:crux}) to generate some self--consistent
model galaxies. First we consider a Hernquist--like DF ($\alpha=1$,
$\gamma=4$) as a suitable model for a cuspy dark--matter halo. We then compute the cored 
equivalent to this model ($\alpha = 0$, $\gamma = 4$). Finally, we investigate a
Jaffe--like model ($\alpha=2$, $\gamma=4$) to represent an elliptical
galaxy.

To investigate these DFs, we implemented the algorithms of Sections
\ref{sec:selfcons} and \ref{sec:anistune} in Python. Given a radial
grid and an initial guess potential $\Phi _0$, our code will provide a
self--consistent model with the spatial and kinematic properties
readily evaluated. The initial guess potential used in each case is the potential 
of the target double power--law density distribution (see Appendix \ref{sec:cuspnormapp}). 
At each iteration, we make a grid of the radial
action as a function of binding energy and angular momentum,
$J_r(E,L)$. Between gridpoints, we use cubic spline interpolation to
find $J_r$. Beyond the end of the energy grid (low binding energies)
we use the Keplerian approximation for the radial action from Equation
(\ref{eq:kepler}):
\begin{equation}
J_r(E<E_\mathrm{min}) \simeq \dfrac{GM}{\sqrt{2E}} - L
\end{equation}
since we are only considering finite--mass models. Once the radial
action grid has been calculated, we numerically integrate the DF to
find the new density and potential. After the first iteration, the
potential is purely numerical, and is only explicitly calculated at
the radial grid--points. We again use cubic spline interpolation to
evaluate the potential between grid--points. Beyond the final
grid--point, we extrapolate the potential in a Keplerian fashion
\begin{equation}
\Phi(r>r_\mathrm{max}) = \Phi(r_\mathrm{max})\dfrac{r_\mathrm{max}}{r}.
\end{equation}
Our condition for convergence is that the change in the potential must
be $< 1\%$ at all the radial gridpoints. To speed convergence, we use
the trick employed by Binney (1982, 2014) \nocite{Bi82}
\begin{equation}
\Phi _{i+1} = (1+\kappa)\Phi _{i,1/2} - \kappa \Phi _i
\end{equation}
where $\Phi _{i,1/2}$ is the potential computed by solving Poisson's
equation for the new density profile and $\Phi _i$ is the potential
from the previous iteration. We find $\kappa = 0.5$ is a reasonable
value here, and our code converges in $\sim 3$ iterations. In order to
test this code, we provided $f(\acts)$ for the isochrone
model~\citep{He59}, which is known entirely analytically, and set
$\Phi _0$ to the \citet{Pl11} potential. Upon convergence, the code
recovers the isochrone potential to great accuracy, with the largest
error $\sim 1\%$ in the very center of the model. This is not
surprising, since our radial grid cannot extend to zero and so the
integration to find the potential at the innermost grid--point must
rely on extrapolation.

To implement the algorithm of Section \ref{sec:anistune} we used
Brent's method (e.g. \citealt{Pr07}) to fix the anisotropies at the
innermost and outermost grid--points on our radial grid, and the
action scale of the anisotropy $J_\beta$.  This is done after the
self--consistent potential and density of the model are found using
$D_0$ from Equation (\ref{eq:consts}), $D_1=1$ and $J_\beta=J_0$. The
DF is then slightly different once $J_\beta$, $D_0$ and $D_1$ are
fixed, even after the transformations of Equations (\ref{eq:farfieldnorm}) and (\ref{eq:centernorm}) 
have been applied. If the potential has changed by more than 1\% at any of our gridpoints, it is iterated 
again until our convergence criterion is met. Typically this takes 1 or 2 iterations.

\begin{figure*}
\begin{centering}
\includegraphics[width=3in]{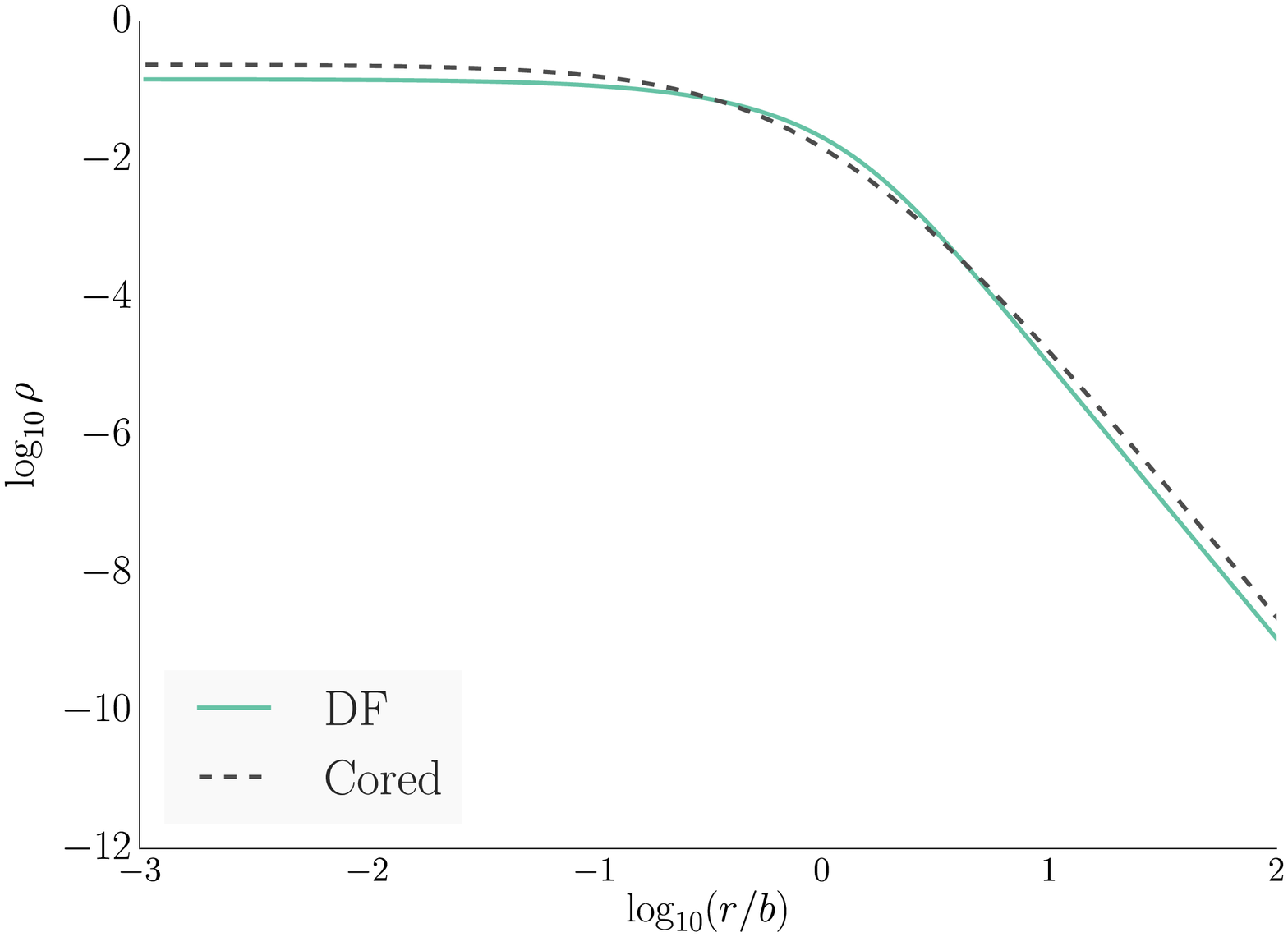}\quad\includegraphics[width=3in]{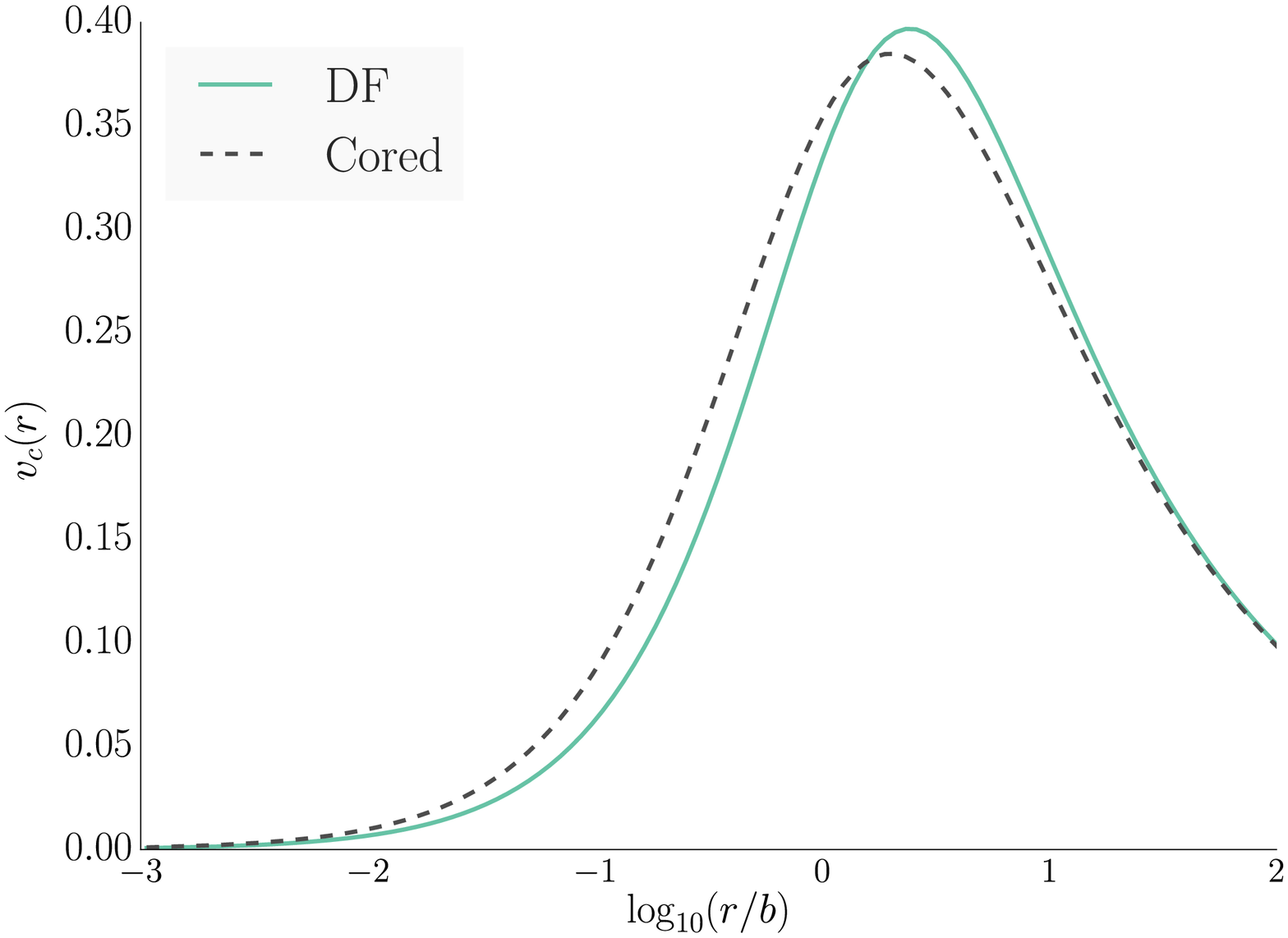} \\
\includegraphics[width=3in]{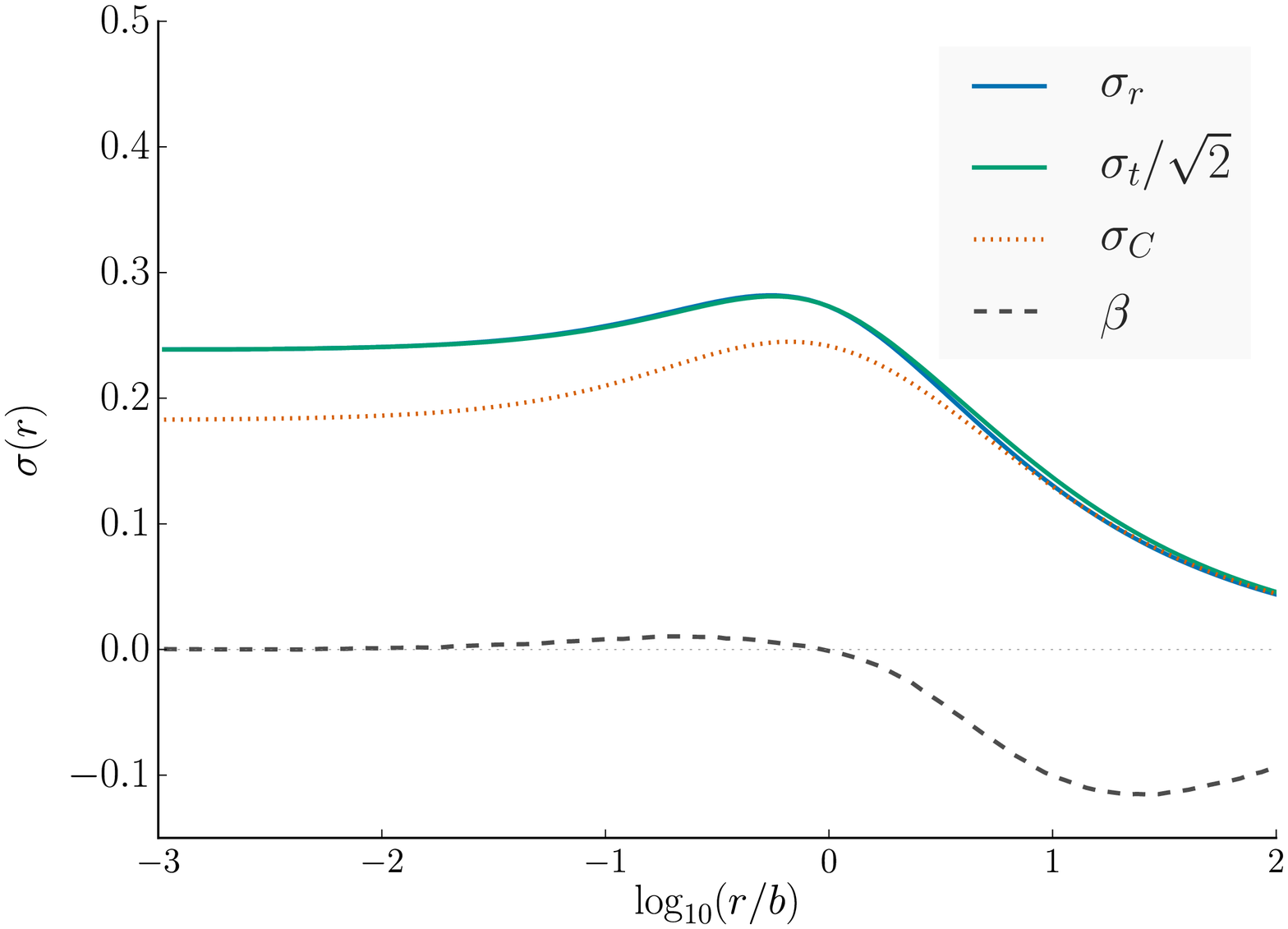}\quad\includegraphics[width=3in]{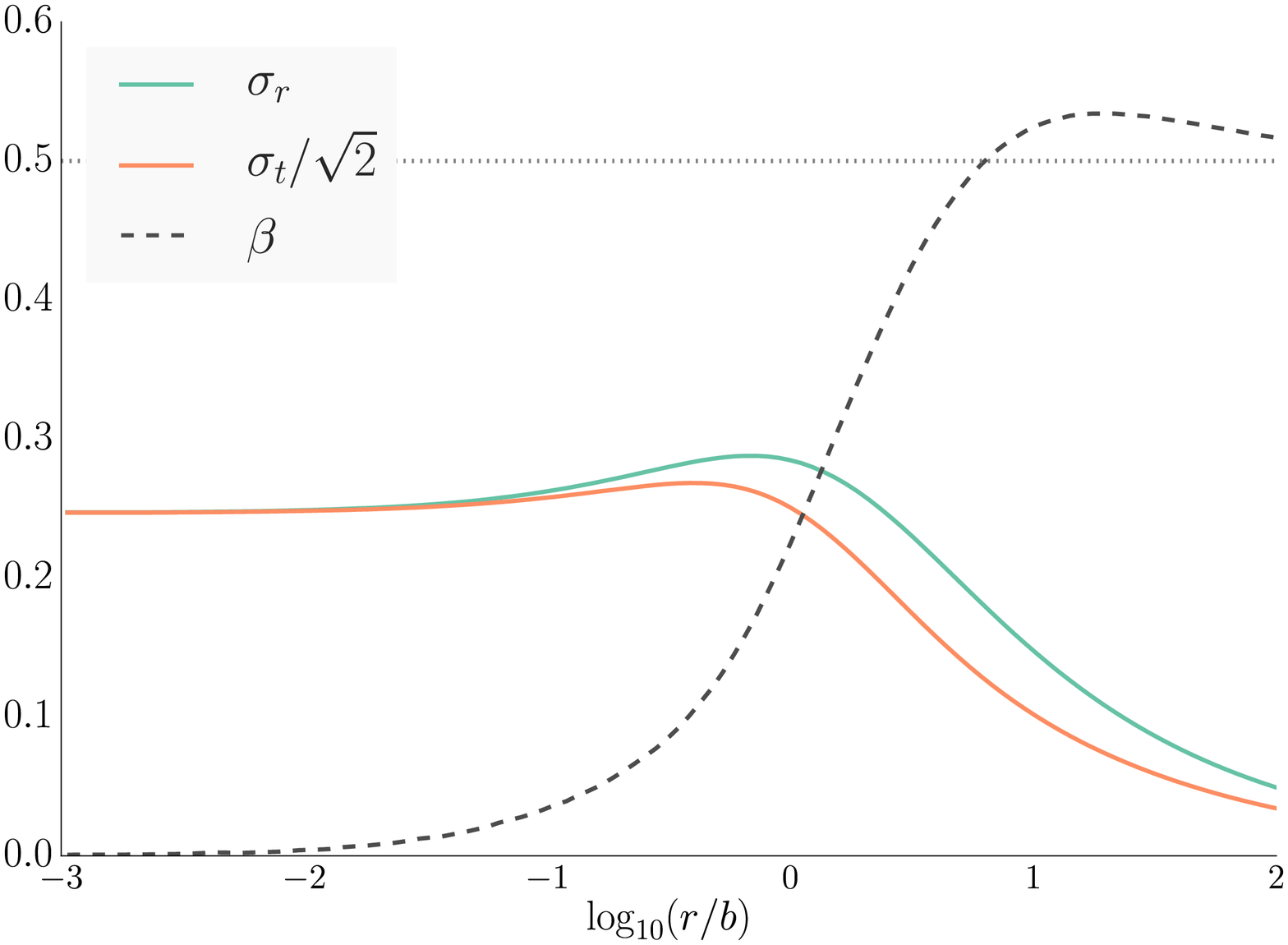}
\caption{Top row: comparison of the density (left) and circular speed
  (right) profiles generated by the DF (\ref{eq:coreDF}) and the cored
  model. The densities are in good agreement, though less so than for
  the other models considered here. The DF appears to produce a
  sharper break than the target density profile, and this is reflected
  in the circular speed curves. Bottom row: the kinematics of the two
  cored dark matter models. The isotropic model (left) becomes very
  mildly tangentially distended with $\beta \sim -0.1$. The dotted line depicts the 
  velocity dispersion profile of the equivalent isotropic cored model ($\sigma _C$), which is in 
  good agreement with our DF beyond $r=b$, but offset in the core of the model. The radially
  distended model (right) switches smoothly from $\beta _0 = 0$ to $\beta _1
  = 0.5$ across a scale of $r_\beta = b$. We set $G=M=b=1$ for these
  models.}
\label{fig:corefig}
\par\end{centering}
\end{figure*}
\begin{figure*}
\begin{centering}
\includegraphics[width=3in]{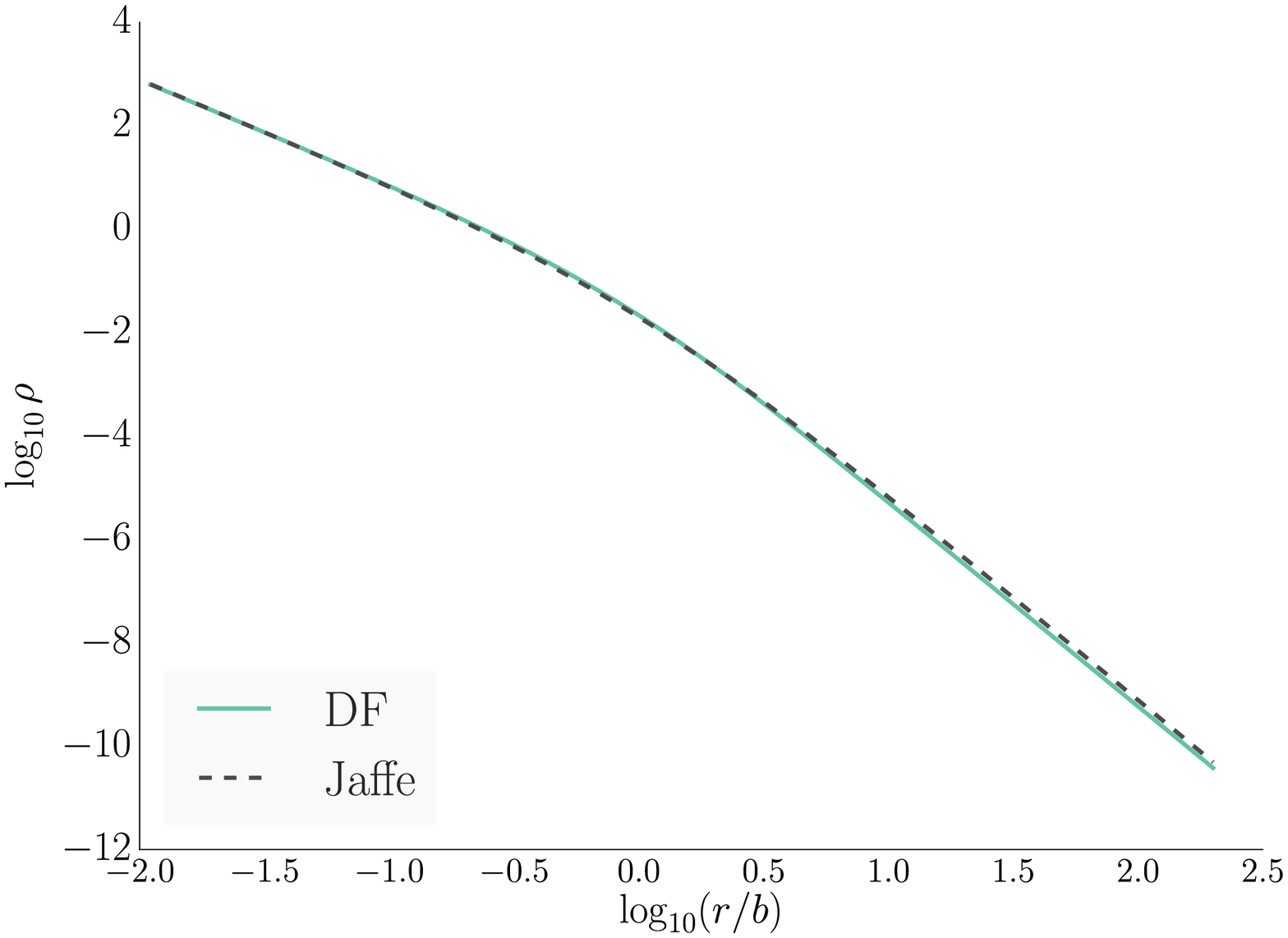}\quad\includegraphics[width=3in]{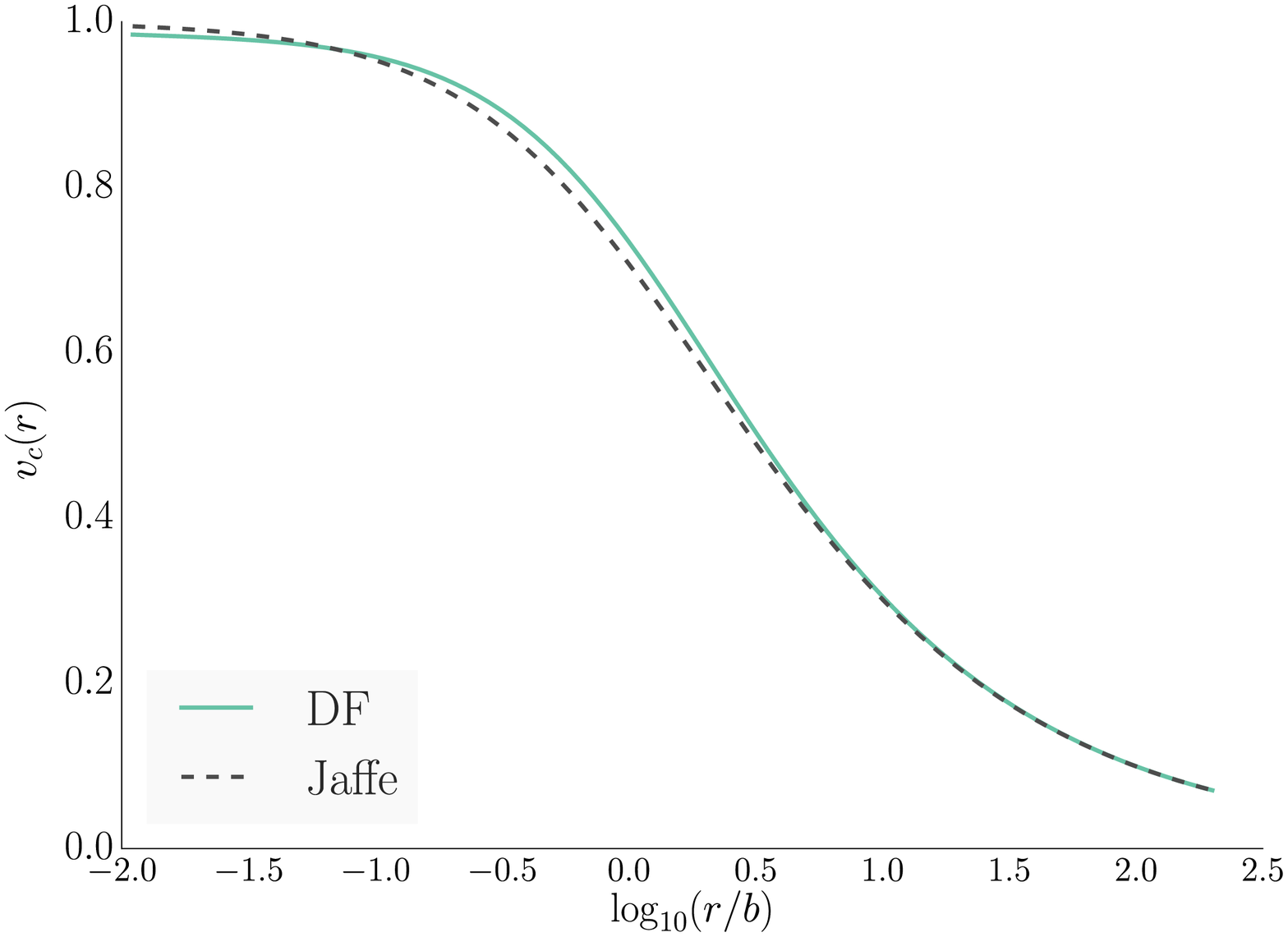} \\
\includegraphics[width=3in]{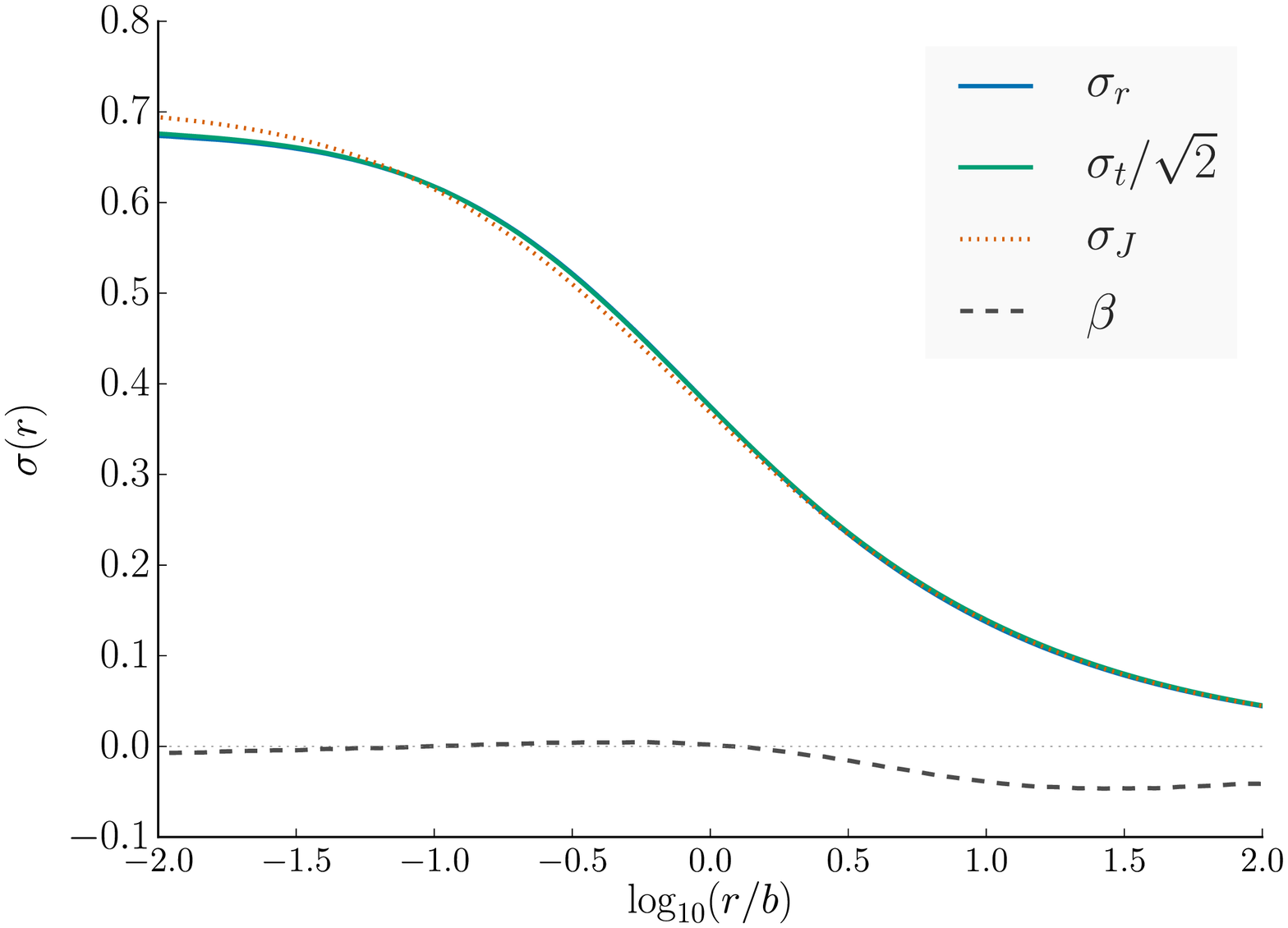}\quad\includegraphics[width=3in]{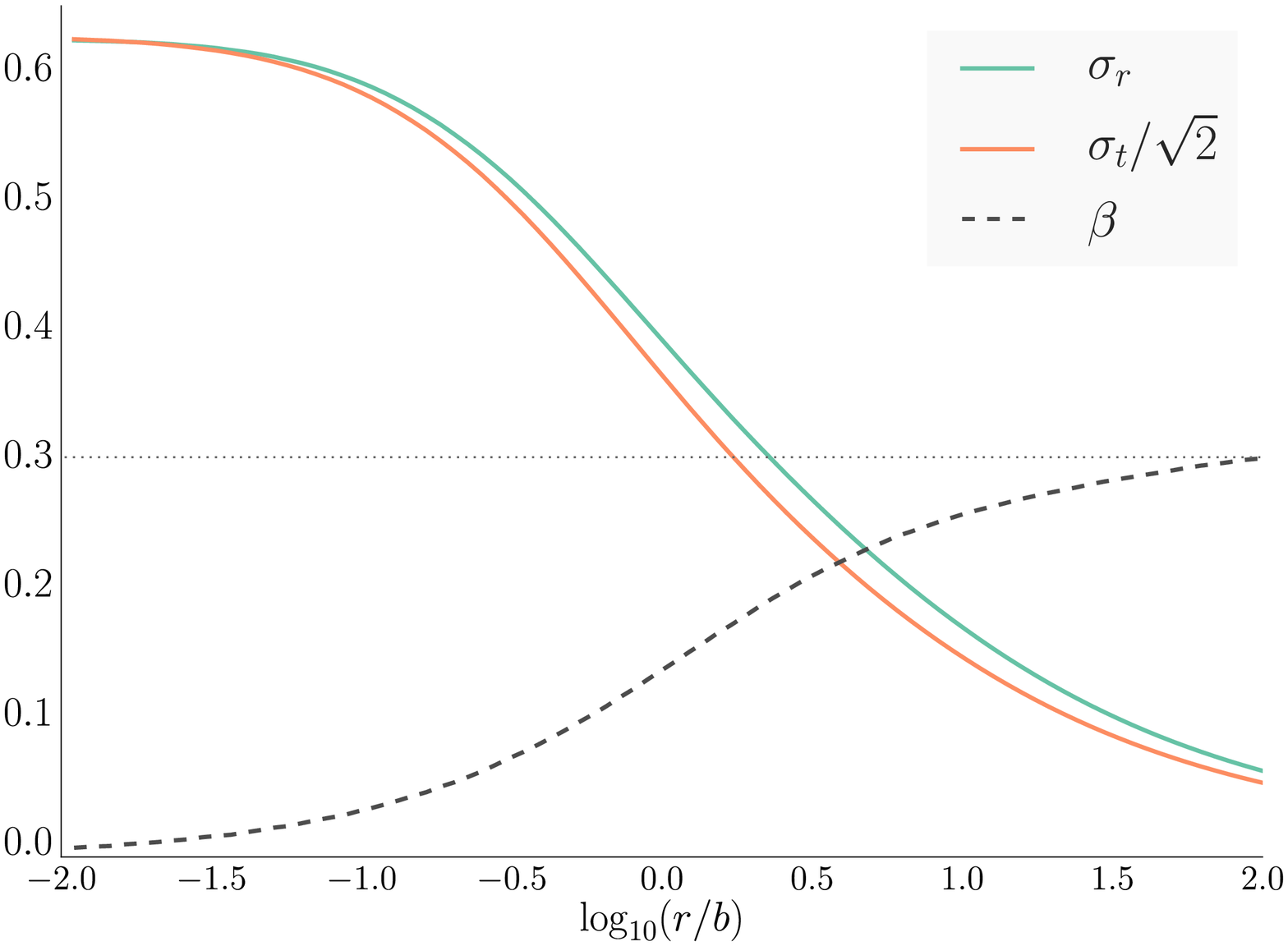}
\caption{Top row: comparison of the density (left) and circular speed
  (right) profiles generated by the DF (\ref{eq:jaffeDF}) and the
  Jaffe model. Both the density and the circular speed are in very
  good agreement. Bottom row: the kinematics of the two elliptical
  galaxy models. The model intended to be isotropic (left) becomes
  very mildly tangentially anisotropic in the far--field, with $\beta \sim
  -0.05$. The dotted line depicts the velocity dispersion profile of the equivalent 
  Jaffe model ($\sigma _J$), which matches our model satisfyingly well.
  The radially anisotropic model (right) switches from $\beta _0 = 0$
  to $\beta _1 = 0.3$ across a scale of $r_\beta = b$. We set
  $G=M=b=1$ for these models.}
\label{fig:jaffefig}
\par\end{centering}
\end{figure*} 

\subsection{Dark--Matter DFs}

Here, we investigate two appropriate DFs for dark matter haloes. One
is cuspy, the other cored. In both cases, we first investigate an
isotropic model and then construct a model with more complicated
kinematics using anisotropy tuning.

\subsubsection{Cuspy Dark Halo}

Cuspy dark matter haloes are often modelled using the \citet{He90}
sphere (e.g. \citealt{Ja01}), a double power--law with $\alpha = 1$
and $\gamma = 4$. In this case, our DF becomes
\begin{equation}
f(\acts) =\dfrac{\mathcal{N} \, M J_0^2}{(2\mathrm{\pi})^3} \,\dfrac{ T(\acts) \, \mathcal{L}(\acts)^{-5/3} }{\left[J_0^2 + \mathcal{L}(\acts)^2\right]^{5/3}}.
\label{eq:hernDF}
\end{equation}
We first consider an isotropic model, where the variable normalisation
factor is calculated to be (see Appendix \ref{sec:cuspnormapp})
\begin{equation}
T(\acts) = \dfrac{0.378 + |\acts|/J_0}{1+|\acts|/J_0}
\end{equation}
and the weighting factor for the actions is
\begin{equation}
D(\acts) = \dfrac{\mathrm{\pi}/\sqrt{3} + |\acts|/J_\beta}{1 + |\acts|/J_\beta}
\end{equation}
After solving for the self--consistent model, we optimized $J_\beta$
using our anisotropy tuning algorithm and found $J_\beta = 0.41
J_0$. The two upper panels of Figure \ref{fig:hernfig} depict the
comparison of the radial density and circular speed profiles of this
model with the Hernquist model of equivalent mass and
scale--length. We find that the DF is in excellent agreement with the
Hernquist model, the only noticeable discrepancy being a slight offset
in the circular speed at small radii. The lower--left panel of Figure
\ref{fig:hernfig} depicts the kinematic properties of this model after
anisotropy tuning has taken place. Our intention, in this case, was to
produce a model as close to isotropic as possible. We can see that
this target has been well--met, with $\beta$ only fluctuating on
scales $\sim 0.01$: these fluctuations cannot be seen by inspection of
the velocity dispersion profiles.

In numerical simulations, it is usually found that dark--matter haloes
are isotropic in the centre and significantly radially anisotropic at
large radii (e.g., \citealt{Ha06,De11b}; although \citet{Wo13b}
demonstrate that this finding is not universal).  As a result, we look
to create a cosmologically realistic DF by using anisotropy tuning to
create a Hernquist--like model that has $\beta = 0$ in the central
regions and $\beta = 0.5$ in the outer parts. The density and circular
speed profiles look identical to the upper two panels of Figure
\ref{fig:hernfig}, and the kinematics are shown in the bottom--right
panel of the same figure. We can see that anisotropy tuning has
worked very nicely, the model moves smoothly between the two values of
$\beta$, only very slightly exceeding $0.5$ in the outer parts. After
tuning, we have
\begin{equation}
T(\acts) = \dfrac{1.18 + |\acts|/J_0}{1+|\acts|/J_0}
\end{equation}
and 
\begin{equation}
D(\acts) = \dfrac{\mathrm{\pi}/\sqrt{3} + 0.59|\acts|/J_\beta}{1 + |\acts|/J_\beta},
\end{equation}
with $J_\beta = 0.19J_0$. We have thus found a very simple DF that
well--represents the phase space structure of dark--matter haloes
found in cosmological simulations.

\subsubsection{Cored Dark Halo}
\label{sec:coreddark}

The density profile at the centre of dark matter haloes is a widely
disputed issue. Although the classical simulations \citep*{Na96}
produce $r^{-1}$ cusps, it is now believed that many haloes could have
constant densities in the center as a result of non--adiabatic
physical processes like active AGN or supernova feedback (e.g.,
\citealt{Go10, Te11, Po12}). In the case of dwarf spheroidal galaxies,
many studies have been carried out in order to determine the nature of
the dark matter density law. Some results favour cores (e.g.,
\citealt{Wa11,Am12,Ag12}), while others offer a different view (e.g.,
\citealt{Br13,St14,Ri14}). As such, it is clear that dynamical models
of both cored and cuspy dark matter haloes are required to resolve the
controversy.

Using our DF (\ref{eq:crux}), we can create isothermal cored profiles
by simply setting $\lambda = 0$. However, here we choose to construct
a ``cored Hernquist" with a non-isothermal core ($\alpha = 0$, $\gamma
= 4$). In this case, one can demonstrate that $\lambda=1$ rather than
$3/2$ as Equation (\ref{eq:asymDF}) would suggest. This is due to a
discontinuity in the behaviour of the ergodic DF of the double-power
law or Gamma models \citep{De93,Tr94} as the inner density slope $\alpha
\rightarrow 0$.  Our DF is then
\begin{equation}
f(\acts) =\dfrac{\mathcal{N} \, M J_0^2}{(2\mathrm{\pi})^3} \,\dfrac{ T(\acts) \, \mathcal{L}(\acts)^{-1} }{\left[J_0^2 + \mathcal{L}(\acts)^2\right]^{2}}.
\label{eq:coreDF}
\end{equation}
For an isotropic model, the variable normalisation and weighting
factor for the actions are given by
\begin{equation}
T(\acts) = \dfrac{5/4 + |\acts|/J_0}{1+|\acts|/J_0}, \quad D(\acts) = \dfrac{2 + |\acts|/J_\beta}{1+|\acts|/J_\beta}.  
\end{equation}
Anisotropy tuning gives $J_\beta = 0.8J_0$. The upper two panels of
Figure \ref{fig:corefig} depict the comparison between the target
density profile and the one produced by our DF. We can see that,
although the behaviour is qualitatively the same, the DF produces a
sharper break in density than we see in the target profile. These
issues could be remedied by altering the strength of the break in
action--space of the DF. Once again, anisotropy tuning is successful
in creating a very nearly isotropic model, with very slight tangential
bias at larger radii (bottom left panel). We then opted to create a
halo with the same anisotropy profile as the cuspy halo model (bottom
right panel), with very similar results. In this case, the variable
normalisation and weighting factor are
\begin{equation}
T(\acts) = \dfrac{3.86 + |\acts|/J_0}{1+|\acts|/J_0}, \qquad
D(\acts) = \dfrac{2 + 0.59|\acts|/J_\beta}{1+|\acts|/J_\beta}
\end{equation}
and $J_\beta = 0.16J_0$.

\section{Elliptical Galaxy DF}

We now perform a similar exercise, but for an elliptical galaxy
DF. The \citet{Ja83} model is widely used to fit the light
distributions of elliptical galaxies, and so we look to construct a DF
that can represent it. The Jaffe model is a double power--law with
$\alpha = 2$ and $\gamma = 4$. We study the DF
\begin{equation}
f(\acts) =\dfrac{\mathcal{N} \, M J_0^2}{(2\mathrm{\pi})^3} \,\dfrac{ T(\acts) \, \mathcal{L}(\acts)^{-2} }{\left[J_0^2 + \mathcal{L}(\acts)^2\right]^{3/2}}.
\label{eq:jaffeDF}
\end{equation}
Once again, we shall first consider an isotropic model followed by a
model with more complex kinematics that are closer to the observed
properties of ellipticals. In the isotropic case, we have
\begin{equation}
T(\acts) = 1
\end{equation}
in other words, the cusp and the far--field are equally weighted. We
then have
\begin{equation}
D(\acts) = \dfrac{\sqrt{\dfrac{2\mathrm{\pi}}{e}} + |\acts|/J_\beta}{1 + |\acts|/J_\beta}.
\end{equation}
Anisotropy tuning finds the optimal value of $J_\beta$ to be
$0.69J_0$. The upper two panels of Figure \ref{fig:jaffefig} compare
the Jaffe model density and circular speed profiles with our DF. Once
again, we can see that the DF reproduces the target model very
nicely. The bottom--left panel depicts the kinematics of this model:
the anisotropy parameter is again minimised effectively by our
algorithm, with the model becoming very mildly tangentially distended
at larger radii, with $\beta \sim -0.05$ at worst.

\begin{figure}
\includegraphics[width=3.4in]{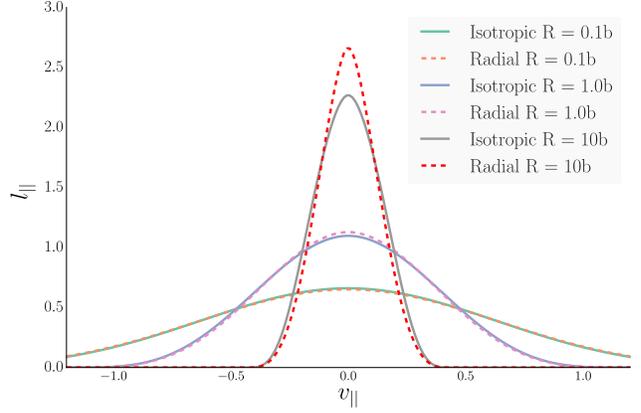}
\caption{Line--profiles for the two elliptical galaxy models
  considered. Full lines correspond to the isotropic model and dashed
  lines are the radially biased model. For each model, the
  line--profile is plotted at $R=0.1b$, $R=b$ and $R=10b$. We can see
  that the profiles are indistinguishable at small radii where the
  models are essentially identical, but then the radially biased model
  has a narrower line--profile at large radii. We have set $G=M=b=1$ for these models.}
\label{fig:lineprofiles}
\end{figure}

As with our dark--matter DF, we now create a model with a more realistic
velocity distributions. Elliptical galaxies are thought to be
isotropic in the central regions, and mildly radially anisotropic in
the outer parts \citep{Kr00}. To mimic this, we build a model with
$\beta = 0$ in the central parts and $\beta = 0.3$ in the outer
parts. The bottom--right panel of Figure \ref{fig:jaffefig} shows the
kinematic properties of this model. Once again, anisotropy tuning has
been successful in creating a model with desirable kinematic
properties. The anisotropic model has weighting factor
\begin{equation}
D(\acts) = \dfrac{\sqrt{\dfrac{2\mathrm{\pi}}{e}} + 0.74|\acts|/J_\beta}{1 + |\acts|/J_\beta}
\end{equation}
and variable normalisation
\begin{equation}
T(\acts) = \dfrac{2 + |\acts|/J_0}{1+|\acts|/J_0}.
\end{equation}
Since elliptical galaxies are seen in projection, a kinematic quantity
of interest is the line--of--sight velocity profile
(line--profile). We can extract this distribution from a DF by
integrating over the line--of--sight and the tangential velocity
components \citep{Ev94}. Let $v_{||}$ be the line--of--sight velocity,
$R$ the projected radial position on the sky and $z$ the
line--of--sight distance. Then
\begin{equation}
L_{||}(R) = \int _{z_{-}} ^{z_{+}} \intd z \int _{-(-2\Phi-v_{||}^2)^{1/2}} ^{(-2\Phi-v_{||}^2)^{1/2}} \intd v_x \int _{-(-2\Phi-v_{||}^2-v_x^2)^{1/2}} ^{(-2\Phi-v_{||}^2-v_x^2)^{1/2}} \intd v_y \quad f(L,J_r)
\label{eq:losvdist}
\end{equation}
where the limits $z_{\pm}$ arise from the finite mass of the model, so
that a particle with energy $E$ is bounded in position via $v_{||}^2 <
-\Phi(R,z)$. In this instance, we shall consider the normalised
line--profile, which is given by
\begin{equation}
l_{||}(R) = \dfrac{L_{||}(R)}{I(R)}
\end{equation}
where $I(R)$ is the surface brightness of the galaxy at projected
radius $R$. Figure \ref{fig:lineprofiles} depicts the line--profiles
for the two models considered here at three radii, $R=0.1b$, $R=b$ and
$R=10b$. The models are essentially identical at small radii, so the
line--profiles are almost indistinguishable. At $R=10b$ however, where
the radial model has $\beta = 0.2$, one can see that the profiles are
notably different: the radial model has a more strongly peaked
line--profile than the isotropic model. This demonstrates the
versatility of these models when fitting to observational
data.

\section{Dwarf Galaxy DF}

Here, we describe another useful DF that can be used to describe the
stellar component of dwarf galaxies or globular clusters, and is
designed to generate a model close to the \citet{Pl11} sphere. We
first derive the DF, and then provide an application in which we relax
the dwarf galaxy DF within one of our dark matter (Hernquist--like)
models.

\subsection{Derivation}

\begin{figure*}
\begin{centering}
\includegraphics[width=3in]{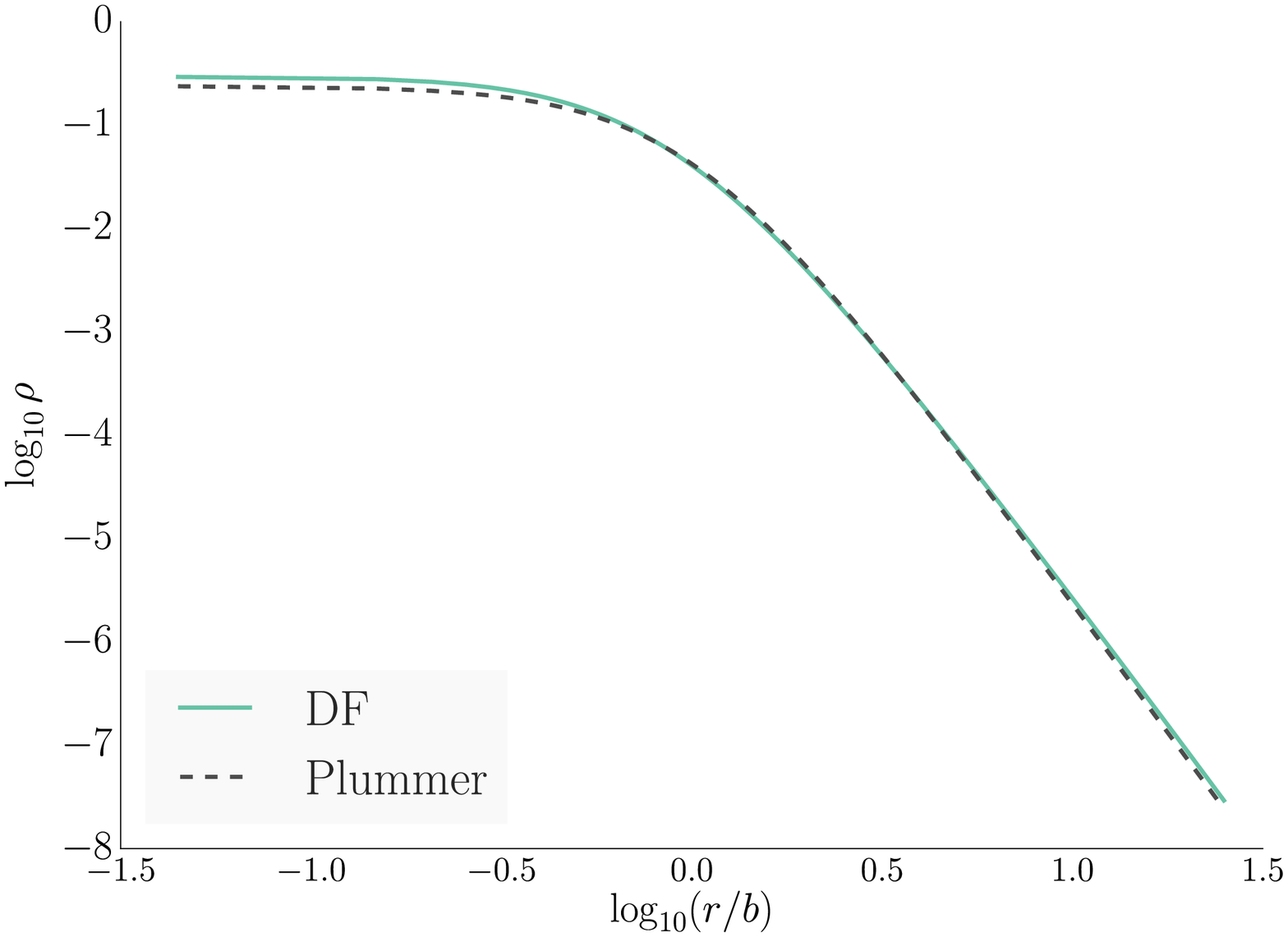}\quad\includegraphics[width=3in]{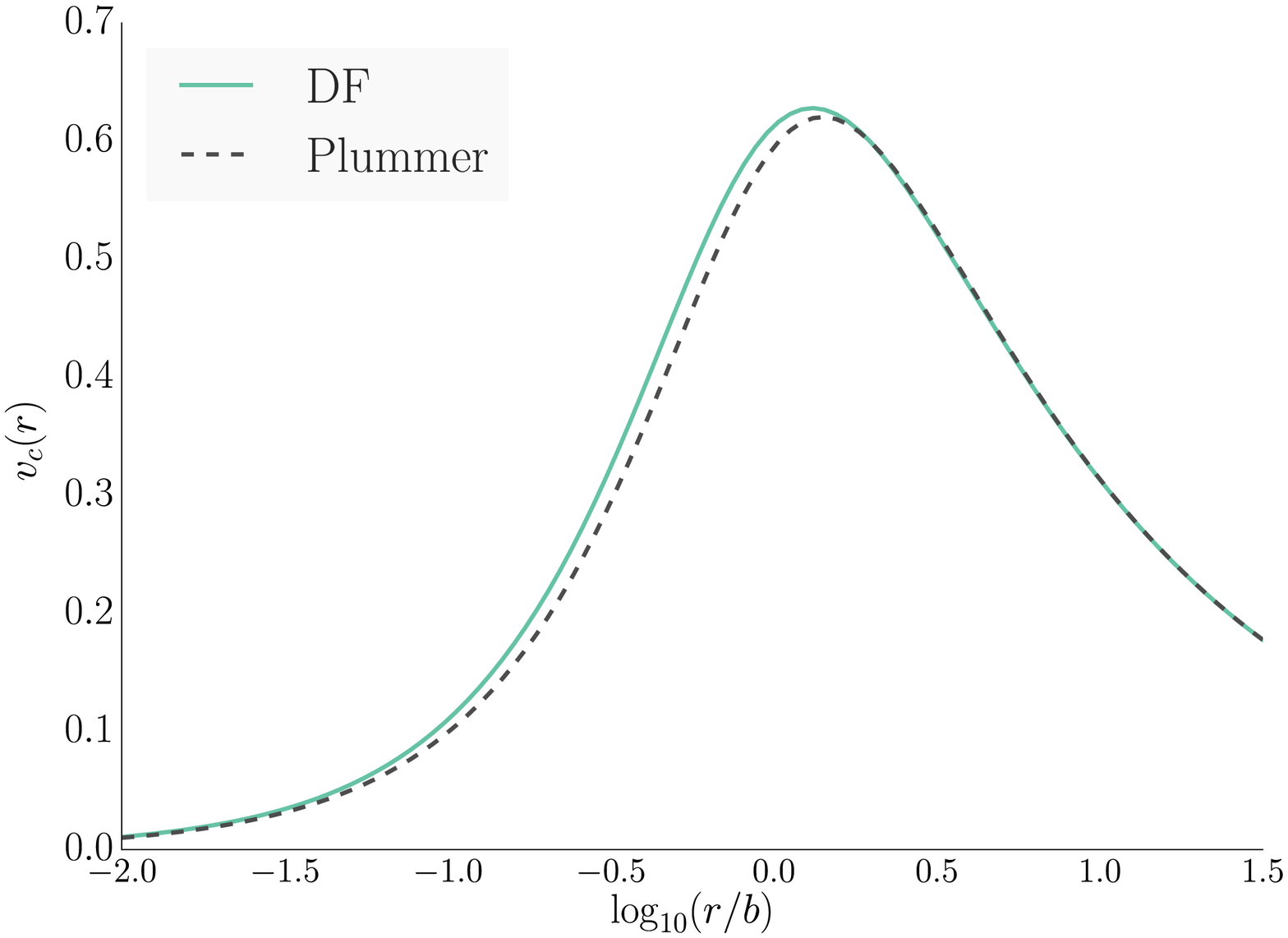}\\ \includegraphics[width=3in]{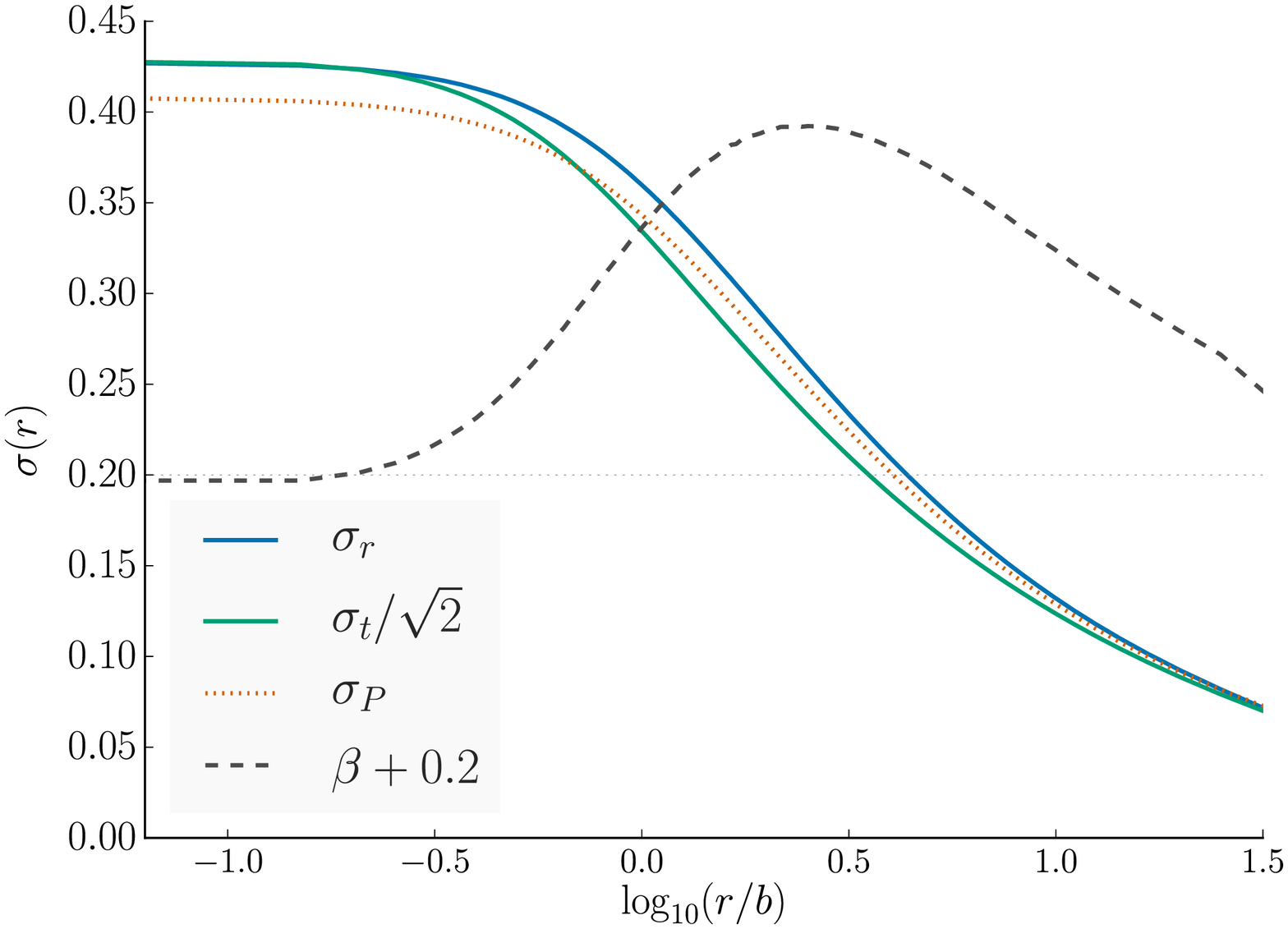}
\caption{Top row: Comparison of the density profile (left) and
  circular speed generated (right) by the DF of Equation
  (\ref{eq:plummerDF}) and the true Plummer model. Bottom: kinematics
  of the model, we see that the anisotropy parameter peaks at
  $\sim0.2$ when $r\sim3b$. The dotted line depicts the velocity dispersion profile of an isotropic Plummer model ($\sigma _P$). We set $G=M=b=1$ for these models.}
\label{fig:plumplot}
\par\end{centering}
\end{figure*}

The \citet{Pl11} sphere has gravitational potential and density
profiles
\begin{equation}
\Phi(r) = \dfrac{-GM}{\sqrt{r^2 + b^2}}\quad;\quad\rho(r) = \left(\dfrac{3M}{4\mathrm{\pi}b^3}\right)\left[1 + \left(\frac{r}{b}\right)^2\right]^{-5/2}.
\label{eq:plummer}
\end{equation}
The density profile is flat in the centre, then declines as $r^{-5}$
at large radii. Naively, then, one might think that the DF of Equation
(\ref{eq:crux}) could be used because the density has two power--law
regimes. However, the Plummer sphere is a polytrope with a very simple
ergodic DF
\begin{equation}
f(E) \propto E^{7/2},
\end{equation}
and so the logic of Section \ref{sec:powerlaws} no longer applies. As
a result, we choose to use a method closer to that employed by
\citet{Ev14}, and construct an approximate Hamiltonian for the Plummer
sphere. Consider the Hamiltonian of an isochrone model with mass $M$
and scale--length $b$:
\begin{equation}
H_\mathrm{iso}(\acts) = \dfrac{-(GM)^2}{2\left[J_r+\frac{1}{2}\left(L+\sqrt{L^2+4GMb}\right)\right]^2}.
\end{equation}
This Hamiltonian varies from the harmonic oscillator when
$|\acts|\ll\sqrt{GMb}$ to Keplerian for $|\acts|\gg\sqrt{GMb}$. The
Plummer potential of Equation (\ref{eq:plummer}) shares the same
functional limits with the isochrone. For this reason, we choose to
use $H_\mathrm{iso}$ as a template for an approximate Plummer
Hamiltonian. The selected ansatz is
\begin{equation}
\mathcal{H}(\acts) = \dfrac{-(GM)^2}{\left[g(J_r)J_r+\frac{1}{2}\left(L+\sqrt{\delta^2 L^2+4GMb}\right)\right]^2},
\end{equation}
where the function $g(J_r)$ is necessary because the coefficient of
$J_r$ changes between the two limiting cases, which is not the case
for the isochrone. We can analytically solve for the constant $\delta$
and easily select a simple function $g(J_r)$ to ensure that this
Hamiltonian coincides with the correct limits at small and large
action. We find;
\begin{eqnarray}
\delta &=& 4\sqrt{2} - 2, \nonumber \\
g(J_r) &=& \dfrac{\sqrt{2}J_r+\sqrt{GMb}}{J_r+\sqrt{GMb}}.
\end{eqnarray}
Now that an approximate Hamiltonian has been constructed, we can
substitute into the ergodic DF of the Plummer sphere to find the
approximation
\begin{equation}
f(\acts) = \dfrac{3\mathcal{N}2^{7/2}G^2M^3b^2}{7\pi^3}\left[\frac{1}{2}\left(L+\sqrt{\delta^2L^2+4GMb}\right)+g(J_r)J_r\right]^{-7}.
\label{eq:plummerDF}
\end{equation}
Figure \ref{fig:plumplot} demonstrates the effectiveness of this
DF. The density and circular--speed profiles of the model are in good
agreement with the true Plummer model. The model becomes moderately
radially anisotropic, where $\beta\simeq0.2$ at $r\sim3b$.

\subsection{Building a Dwarf Galaxy}

\begin{figure}
\includegraphics[width=3.4in]{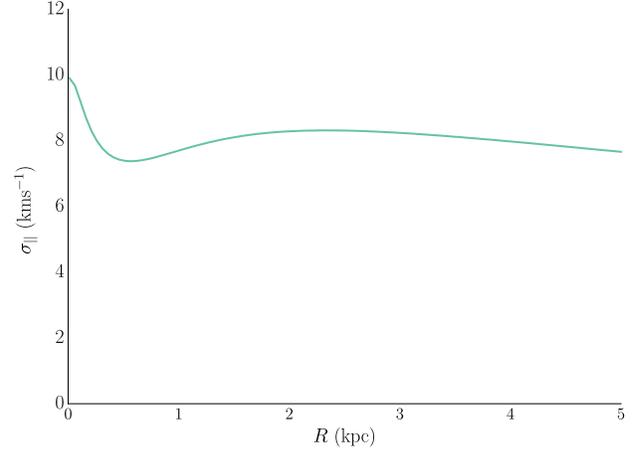}
\caption{The line--of--sight velocity dispersion profile of the
  stellar matter in our dwarf galaxy model within $5\kpc$. The profile
  is essentially flat, as is typical in observations of dwarf
  spheroidals, though there is a slight upturn in the profile as $R
  \rightarrow 0$.}
\label{fig:dwarflos}
\end{figure}

We now briefly demonstrate the usefulness of $f(\acts)$ models by
relaxing a dwarf galaxy DF and a dark matter DF simultaneously, so
that the full DF is given by
\begin{equation}
f(\acts) = f_{\mathrm{dm}}(\acts) + f_*(\acts).
\end{equation}
$f_*$ is given by the DF of Equation (\ref{eq:plummerDF}) and we
choose $f_{\mathrm{dm}}$ to be the cored double power--law with $\alpha
= 0$ and $\gamma = 4$ from Section \ref{sec:coreddark}. We choose the parameters of our
dwarf--galaxy to be
\begin{equation}
M_\mathrm{dm} = 10^9 \Ms \, , M_* = 10^7 \Ms \, , a_\mathrm{dm} = 1 \kpc \, , a_* = 300 \pc.
\end{equation}
where $M_\mathrm{dm}$ and $M_*$ are the dark matter and stellar masses
respectively, whilst $a_\mathrm{dm}$ and $a_*$ are the scalelengths.
This leads to two natural action scales in the model, which are
$J_\mathrm{dm} = \sqrt{GM_\mathrm{dm}a_\mathrm{dm}}$ and $J_* =
\sqrt{GM_*a_*}$. After constructing the total DF of the model, we used
the iterative procedure from Section \ref{sec:selfcons} to compute the
self--consistant gravitational potential produced by this model. After
this calculation is complete, one can compute properties of either
component of the model by simply performing integrals over that part
of the DF alone. For example, the density of the stellar component of
our model is simply given by
\begin{equation}
\rho _*(r) = \int \intd^3 \boldsymbol{v} \, f_*(\acts).
\end{equation}
In dwarf galaxies, the line--of--sight velocity dispersion, $\sigma
_{||}(R)$, of the stars is one of the only kinematic quantities
available via observations. To compute this in our model, we evaluate
the integral
\begin{equation}
\sigma _{||}^2(R) = \dfrac{1}{I_*(R)}\int \intd z \int \intd v_x \int \intd v_y \int \intd v_{||} \quad v_{||}^2 f_*(\acts),
\end{equation}
where again $R$ is the projected radius, $z$ is the line--of--sight
distance and $v_x$, $v_y$ are the tangential components of
velocity. As in Equation (\ref{eq:losvdist}), we integrate over bound
orbits. Figure \ref{fig:dwarflos} depicts this quantity for the model
we produce here. We see that the dark matter has produced a
largely flat profile as is seen in observations~\citep[see
  e.g.,][]{Wa09}, although as $R\rightarrow 0$ the dispersion
increases somewhat. This effect is due to the comparatively modest
central mass-to-light ratio in the model, so this is a reasonable
representation of a large dwarf spheroidal like Fornax rather than its
smaller, overwhelmingly dark matter dominated cousins.  We plan to
return to the problem of dwarf galaxy modelling in action-angle
coordinates in the near future, but this is a simple example of what
can be done with these models.

\section{Conclusions}

Dynamical models constructed in action coordinates have many desirable
properties but, until recently, two issues have stood in the way of
their use. First, it was not known how to compute actions in
aspherical potentials in the general case. This in turn meant that
most astrophysical systems, such as disks or flattened dark matter
haloes, could not be modelled. However, the past couple of years have
seen rapid development in this area. \citet{Bi12a} provided an
approximate method for computing actions in axisymmetric potentials
(refined in \citealt{Bi14}) by locally approximating the potential as
a St\"{a}ckel potential, which was recently generalised to triaxial
potentials by \citet{Sa15}. An alternative method based on deforming
orbital tori by the use of generating functions was also recently
discovered by \citet{Sa14a} and \citet{Bo14}, allowing very accurate
computation of the actions in wide classes of triaxial potentials. 

The second problem was a lack of insight into the form of $f(\acts)$
for different types of galaxies. The isochrone potential is the only
known model with a completely tractable $f(\acts)$
\citep[e.g.,][]{Ev90}, and techniques had not been developed to
construct models suitable for more realistic components of
galaxies. Recently, however, Binney (2010, 2012) provided distribution
functions for galactic disks, and \citet{Po13} suggested an ansatz for
a dark--matter halo distribution function. Nonetheless, few models
existed with which to construct realistic pressure-supported
galaxies. It is this problem that we have addressed in this
work. \citet{Wi14a} provided a method for approximating the
Hamiltonians of power--law potentials, which can be used to construct
Hamiltonians for scaled spherical potentials \citep{Ev14}. In this
paper, we have demonstrated how these approximations can be used to
construct physically realistic distribution functions for spherical
galaxy components. The DFs presented here can also be relatively
easily generalised to become flattened, triaxial or rotating by using
methods such as action scaling \citep{Bi14}. It would seem that now is
an exciting time for this field and these models should prove
invaluable for understanding our galaxy (e.g. Piffl et
al. 2014\nocite{Pi14}) and external systems.

We provided two DFs: one designed to emulate double power--law
profiles and the other to approximate the Plummer sphere. The double
power--law DF has the additional property that one may tune the
anisotropy profile by adjusting three of the parameters in the model
independently. As such, the DF of Equation (\ref{eq:crux}) can provide
a wide variety of models with differing density profiles and
kinematics. We then applied this DF in two ways. First, we constructed
a model designed to mimic the Hernquist profile and demonstrated that
anisotropy tuning is effective at creating a nearly isotropic model,
if desired. The model was then altered to become isotropic in the
central regions, changing to radial anisotropy at large radii --
consistent with the dark matter haloes found in cosmological
simulations. This procedure was carried out for both cored and cusped
dark haloes. Second, we created a Jaffe--like model to represent an
elliptical galaxy. Once again, anisotropy tuning was very effective in
creating a very nearly isotropic model. Subsequently, we again changed
the kinematics of the model so that it became mildly radially
anisotropic in the far--field. As an example of how these models might
be applied to data, we computed and compared the line--profiles of the
two elliptical galaxy models. Our final section described the
derivation of a DF for the stellar component of dwarf galaxies or
globular clusters. This DF is derived by explicitly approximating the
Hamiltonian of the model, another promising method for constructing
$f(\acts)$ models.

A difficulty with our ansatz for the double power--law DF is that it
struggles to replicate models with shallow cusps ($\alpha \sim 0.5$).
Experimentation shows that the cusp produced is generally too
steep. Although these models are less often used to represent galaxies
and dark haloes, it is nonetheless a defect that our ansatz cannot
reproduce the full physical range of behaviour. Apparently, the
shallow cusped double power--law models do not possess ergodic DFs
that are well--represented by double power--laws in binding
energy. Interestingly, however, our DF does reproduce cored profiles
well. For example, one simply sets $\lambda = 0$ to create a model
with an isothermal core.

During the completion of this work, \citet{Po14} produced a DF closely
related to that of Equation (\ref{eq:crux}). Their models,
however, differ from ours in several ways. They do not use the
Williams, Evans \& Bowden (2014) approximation to the Hamiltonian of
power-law potentials in the construction of their DF, but instead
approximate the equivalent factor to $D_0$ by $2$, corresponding to
the harmonic oscillator potential. Related to this is that their DF
does not contain an equivalent function to $D(\acts)$, which means
that they cannot tune the anisotropy of the models they
produce. Finally, they also do not include the variable normalisation
factor $T(\acts)$ in their DF. Nonetheless, the basic approach of the
two papers is similar, matching results in different power-law regimes 
to build double power--law models.

There are a few different directions in which this work could be
profitably developed. First, these DFs can be flattened, set rotating
or even made triaxial using the previously mentioned approaches
already in the literature. It will be interesting to investigate the
properties of such models, since they are arguably the simplest avenue
available to us for creating self--consistent models of this
kind. Another possibility to be explored is the construction of
DFs that can well--represent more complex models that are commonly
used, such as the Einasto profile. Anisotropy tuning is also a
promising technique, as one can conceive of many DFs of the form
$f\left[L+D(\acts)J_r\right]$, and perhaps a more flexible, general
form for $D(\acts)$ can be found.

\section*{Acknowledgments}
AW is supported by the Science and Technology Facilities Council. We
thank Jason Sanders and Carlo Nipoti for some useful conversations, as
well as the referee for a thorough and useful report.

\bibliography{m2m}
\bibliographystyle{mn2e}

\appendix

\section{Calculating $S_\alpha$ and $S_\gamma$ when $\gamma = 4$}
\label{sec:cuspnormapp}

Here, we give an example of the calculation of the relative
normalisation factors in the double power--law DF. When $\gamma = 4$,
these models are known as the Gamma models~\citep{De93,Tr94}. The
gravitational potential for such a model ($\gamma \neq 2$) is given by
\begin{equation}
\Phi(r) = -\dfrac{GM}{(2-\alpha)b}\left[1-\left(\dfrac{r}{r+b}\right)^{2-\alpha}\right]
\end{equation} 
Following \citet{De93}, we define the following quantities 
\begin{eqnarray}
\Psi &=& -\dfrac{b\Phi}{GM}, \nonumber \\
\varepsilon &=& \dfrac{bE}{GM}, \\
y &=& \left[1 - (2-\alpha)\Psi\right]^{1/(2-\alpha)}, \nonumber
\end{eqnarray}
which means the integral expression for the ergodic DF is 
\begin{equation}
\label{eq:dehnenDF}
f(\varepsilon) = C \int _0 ^\varepsilon \intd \Psi \dfrac{(1-y)^2[\alpha + 2y + (4-\alpha)y^2]}{y^{4-\alpha}\sqrt{\varepsilon - \Psi}},
\end{equation}
where $C$ is a constant. We now wish to expand this expression in the
limits $\varepsilon \rightarrow 0$ (low binding energies) and
$\varepsilon \rightarrow \Psi(0)$ (high binding energies). We restrict
ourselves to the cases $0 < \alpha < 2$. At low binding energies $y
\simeq 1 - \Psi$ and we can expand the integrand to first order as
\begin{equation}
f(\varepsilon \rightarrow 0) = C\int _0^\varepsilon \intd \Psi \dfrac{6\Psi^2}{ \sqrt{\varepsilon - \Psi}}
\end{equation}
giving the result
\begin{equation}
f(\varepsilon \rightarrow 0) = C\dfrac{32}{5}\varepsilon ^{5/2}.
\end{equation}
In this regime, the binding energy is given by the Kepler Hamiltonian
of Equation (\ref{eq:kepler}). Upon substitution this gives
\begin{eqnarray}
f(\varepsilon \rightarrow 0) &=& C\dfrac{32}{2^{5/2}5}(L+J_r)^{-5} \nonumber \\
\implies S_\gamma &=& \dfrac{32C}{2^{5/2}5}.
\end{eqnarray}
We now turn to high binding energies to compute $S_\alpha$. Let
$\Delta = \Psi(0)-\varepsilon$ and $\Psi(0) - \Psi = x$, so that
Equation (\ref{eq:dehnenDF}) is written
\begin{equation}
f(\varepsilon) = C\int \intd x \dfrac{(1-y)^2[\alpha + 2y + (4-\alpha)y^2]}{y^{4-\alpha}\sqrt{x-\Delta}}.
\end{equation}
At high binding energies, the integrand is strongly peaked around $x =
\Delta$. In this region:
\begin{equation}
y \rightarrow \left[(2-\alpha)x\right]^{1/(2-\alpha)} \rightarrow 0,
\end{equation}
Which allows us to expand the integrand, to first order, as 
\begin{equation}
f(\varepsilon \rightarrow \Psi(0)) = \dfrac{C\alpha}{(2-\alpha)^{(4-\alpha)/(2-\alpha)}}\int _\Delta ^{\Psi(0)} \dfrac{\intd x}{x^{(4-\alpha)/(2-\alpha)}\sqrt{x-\Delta}}.
\end{equation}
We then Taylor expand the integral in $\Delta$ to give \citep{Ve14}
\begin{equation}
f(\varepsilon \rightarrow \Psi(0)) = C\dfrac{\sqrt{\mathrm{\pi}}\Gamma\left(\frac{1}{2} - \frac{2}{\alpha - 2}\right)}{\Gamma\left(\frac{\alpha - 4}{\alpha - 2}\right)}\dfrac{\alpha}{(2-\alpha)^{(4-\alpha)/(2-\alpha)}}\Delta^{(\alpha-6)/2(\alpha-2)}.
\end{equation}
To obtain the DF as a function of the actions, we use the WEB
approximation for the Hamiltonian. Using the same definitions for
$\epsilon$ and $\zeta$ as in Section \ref{sec:powerlaws} and setting
$v_0^2 = GM/b$, this is given by
\begin{equation}
\mathcal{H}(\acts) = \dfrac{v_0 ^{2\zeta / \epsilon}}{\zeta b^\zeta} (L + D J_r)^\zeta.
\end{equation}
Upon substitution, this finally gives
\begin{equation}
S_\alpha = C\dfrac{\sqrt{\mathrm{\pi}}\Gamma\left(\frac{1}{2} - \frac{2}{\alpha - 2}\right) \zeta^{(6-\alpha)/2(2-\alpha)}}{\Gamma\left(\frac{\alpha - 4}{\alpha - 2}\right)}\dfrac{\alpha}{(2-\alpha)^{(4-\alpha)/(2-\alpha)}}.
\end{equation}
In practice, we are interested only in the ratio $S_\alpha /
S_\gamma$, because the normalisation factor $\mathcal{N}$ in Equation
(\ref{eq:crux}) takes care of absolute differences. To that end, we
set
\begin{eqnarray}
S_\alpha &\rightarrow& S_\alpha/S_\gamma, \nonumber \\
S_\gamma &\rightarrow& 1. 
\end{eqnarray}
A similar calculation for the case $\alpha=0$ leads to $S_\alpha =
5/4$.

\end{document}